\documentclass[12pt]{article}

\usepackage{graphics,epsfig}
\usepackage{amsbsy}
\usepackage{amsfonts}
\usepackage{amsmath}
\usepackage{graphicx}

\def\id{\protect{{1 \kern-.28em {\rm l}}}}

\def\k{\kappa}

\def\p{{\partial}}
\def\nn{\nonumber}

\def\dalemb#1#2{{\vbox{\hrule height .#2pt
        \hbox{\vrule width.#2pt height#1pt \kern#1pt
                \vrule width.#2pt}
        \hrule height.#2pt}}}

\let\a=\alpha \let\b=\beta \let\g=\gamma \let\d=\delta \let\e=\epsilon
\let\z=\zeta  \let\th=\theta  \let\k=\kappa
\let\l=\lambda \let\m=\mu  \let\x=\xi \let\p=\pi 
\let\s=\sigma \let\t=\tau   \let\c=\chi 
 \let\vep=\varepsilon
\let\w=\omega      \let\G=\Gamma \let\D=\Delta \let\Th=\Theta \let\L=\Lambda
 \let\P=\Pi \let\S=\Sigma  
\let\C=\Chi \let\W=\Omega
\let\la=\label \let\ci=\cite 
  
\def\nn{\nonumber} \def\bd{\begin{document}} \def\ed{\end{document}}
\def\ds{\documentstyle} \let\fr=\frac \let\bl=\bigl \let\br=\bigr
\let\Br=\Bigr \let\Bl=\Bigl
\let\bm=\bibitem
\let\na=\nabla
\def\tU{{\widetilde U}}
\let\pa=\partial \let\ov=\overline
\def\ie{{\it i.e.\ }}
\newcommand{\be}{\begin{equation}}
\newcommand{\ee}{\end{equation}}
\def\ba{\begin{array}}
\def\ea{\end{array}}
\def\ft#1#2{{\textstyle{{\scriptstyle #1}\over {\scriptstyle #2}}}}
\def\fft#1#2{{#1 \over #2}}
\def\F#1#2{{ F_{#1}^{(#2)} }}
\def\cF#1#2{{ {\cal F}_{#1}^{(#2)} }}

\def\={\, =\, }
\def\+{\, +\, }
\def\-{\, -\, }

\def\R{{\bf R}}
\def\sst#1{{\scriptscriptstyle #1}}
\def\oneone{\rlap 1\mkern4mu{\rm l}}
\def\e7{E_{7(+7)}}
\def\td{\tilde}
\def\wtd{\widetilde}
\def\im{{\rm i}}
\newcommand{\ho}[1]{$\, ^{#1}$}
\newcommand{\hoch}[1]{$\, ^{#1}$}
\newcommand{\bea}{\begin{eqnarray}}
\newcommand{\eea}{\end{eqnarray}}
\newcommand{\ra}{\rightarrow}
\newcommand{\lra}{\longrightarrow}
\newcommand{\Lra}{\Leftrightarrow}
\newcommand{\ap}{\alpha^\prime}
\newcommand{\bp}{\tilde \beta^\prime}
\newcommand{\cB}{{\cal B}}
\newcommand{\cO}{{\cal O}}
\newcommand{\vecx}{\vec{x}}
\newcommand{\vecy}{\vec{y}}
\newcommand{\vecp}{\vec{p}}
\newcommand{\vecq}{\vec{q}}
\newcommand{\tr}{{\rm tr} }
\newcommand{\Tr}{{\rm Tr} }

\newcommand{\cL}{{\cal L}}
\newcommand{\cA}{{\cal A}}
\newcommand{\cD}{{\cal D}}
\def\sst#1{{\scriptscriptstyle #1}}

\def\ve{\varepsilon}
\def\vf{\varphi}
\def\F{\Phi}
\def\wg{\wedge}

\def \foot {\footnote}
\def \bi{\bibitem}

\def \tr {{\rm tr}}
\def \ha {{1 \over 2}}
\def \td {\tilde}
\def \ci{\cite}
\def \N {{\mathcal N}}
\def \ww {\Omega}
\def \const {{\rm const}}
\def \ss {\sum_{i=1}^3 }
\def \t {\tau}
\def\S{{\mathcal S} }
\def \nn {\nu}
\def \XX {{\rm X}}

\def \lra {\leftrightarrow}
\def \vom {{\bar \omega}}
\def \E {{\mathcal  E}} \def \J {{\mathcal  J}}
\def \YY {{\rm Y}}

\def \d {\del}
\def \rJ {{J}}
\def \sms {sigma models\ }
\def \sm {sigma model\ }
\def \L {\Lambda}
\def \gl {\ell}
\def \tr {{\rm tr\ }}
\def\z{\zeta}
\def\zi{\zeta_1}
\def\zii{\zeta_2}
\def\K{\mbox{K}}
\def\eE{\mbox{E}}   \def \vt {\vartheta}
\def \vr {\varrho}
\def \wup {w}

\def\dg{\dagger}
\def\a{\alpha}
\def\b{\beta}
\def\e{\varepsilon}
\def\p{\phi}
\def\ap{\alpha^\prime}
\def\I{{\cal I}}

\def\R{{\bf R}}
\def\Z{{\bf Z}}
\def\C{{\bf C}}
\def\P{{\bf P}}
\def\xb{{\bar X}}
\def\Tr{{\rm  Tr}}
\def\tr{{\rm  tr}}

\def \del{\partial}
\def \a {\alpha}
\def \aa {{\a'}}
\def\g{\gamma}
\def\s{\sigma}
\def\z{\zeta}
\def\zi{\zeta_1}
\def\zii{\zeta_2}
\def\ov{\over}

\def\I{{\cal I}}
\def\J{{\mathcal J}}
\def \ok {{1\ov \k}}
\def\LL{{\mathcal L }}
\def \jL {{J}}
\def \om {\omega}
\def \cL {{\mathcal L}} \def \cH {{\mathcal H}}
\def\E{{\mathcal E}}
\def\w{\omega}
\def\b{\beta}
\def\l{\lambda}
\def\eps{\epsilon}
\def\vep{\varepsilon}
\def \De {{\mathcal D}}

 \def \cV {{\cal V}}
\def  \Jt {  {J}_{\rm tot}    }

\def \k {\kappa}
\def\foot{\footnote}
\def \four{{\textstyle {1\ov 4}}}
 \def \third { \textstyle {1\ov 3
}}
\def\det{\hbox{det}}
\def \ci {\cite}

\def \foot {\footnote}
\def \bi{\bibitem}

\def \tr {{\rm tr}}
\def \ha {{1 \over 2}}
\def \tid {\tilde}
\def \vv {{\rm v}}
\def \tl {{\tilde \l}}
\def \XX {{\rm X}}
\def \ta {{\tilde \a}}
\def \fo { {1\ov 4}}
\def \ep {\epsilon}
\def \inti {{\int^{2\pi}_0 {d \sigma \ov 2 \pi}}}

\def \d {\partial}
\def \K {{\rm S}}
\def \el {\ell}
\def \Tr {{\rm Tr}}
\def \P {\Phi}
\def \l  {\lambda}
\def \tl {{\tilde \l}}
\def \bl {{\tilde \l}}
\def \const {{\rm const}}
\def \V {v}

\def \bv {v^*}
\def \vv {{\rm v}}
\def \LL {{\mathcal L}}
\newcommand{\PV}[1]{P_{\!\!_{V_{#1}}}}

\def \bL {\ell}
\def \M {{\mathcal M}}
\def \N {{\mathcal N}}
\def \S {{\rm S}}
\def \vn {\vec n}
\def \tl {\td \l}
\def \td {\tilde}
\def \Prod {\Pi}
\def \O {{\mathcal O}}
\def \Q {{\rm  Q}}
\def \D {\Delta}
\def \N {{\mathcal N}}
\def\tN{{\tilde N}}

\def \m {\mu}
\def \vs {\vec \s}
\def \ie {i.e.}

\def \cD {{\cal D}}

\def  \le  {\l_{\rm eff}}

\def \rS {{\rm S}}
\def\as{{\a}}
\newcommand{\bra}[1]{\mbox{$\langle #1 |$}}
\newcommand{\ket}[1]{\mbox{$| #1 \rangle$}}

\newcommand{\auth}{AUTHORS}

\def\thb{\bar{\theta}}
\def\Thb{\bar{\Theta}}
\def\barp{\bar{p}}
\def\barq{\bar{q}}
\def\barc{\bar{c}}
\def\bard{\bar{d}}
\def\e{\epsilon}

\def \bi{\bibitem}
\def \la {\label}

\def \l {\lambda}
\def\foot{\footnote}
\def \tl  {{\tilde \l}}
\def \sql {{\sqrt \l}}
\def \adss {$AdS_5 \times S^5$\ }
\newcommand{\rf}[1]{(\ref{#1})}
\def \ov {\over}

\def\th{\theta}
\def\Th{\Theta}
\def\vth{\vartheta}

\def\vth{\vartheta}

\def\ra{\rightarrow}
\def\N{{\cal N}}
\def\F{{\cal F}}
\def\cc{\circ}
\def\eqv{\equiv}

\def\ni{\noindent}
\def \ha{{1\ov 2}}
\def \bw {{\rm w}}

\def\r{{\rm r}}

\def \cT {{\cal T}}
\def \no {\nonumber}

\def \J {\mathcal{J}}
\def \del {\partial}

\def \bps {{\bar \psi}}
\def \sqbl {\sqrt{\bar \lambda}}

\def\dF{\dot{F}}
\def\dG{\dot{G}}
\def\df{\dot{f}}
\def \E {{\cal E}}

\def \S {{\cal S}}
\def \J {{\cal J}}

\def\ms{\mathcal{S}}
\def\mj{\mathcal{J}}
\def\soj{\fr{\ms}{\mj}}
\def \R {{\bf R}}
\def \om {\omega}
\def \tH {\widetilde H}
\def \bE {\bar E}
\def \x {{\cal X}}

\def \hV {{\hat V}}

 \def \bb {\bar \beta}
\def \W {{\cal E}}

\def \bi{\bibitem}
\def \la {\label}

\def \l {\lambda}
\def\foot{\footnote}
\def \tl  {{\tilde \l}}
\def \sql {{\sqrt \l}}
\def \sqtl {{\sqrt {\tilde \l}}}

\def \HH {{\rm E}}
\def \cS {{\cal S}}
\def \cL {{\cal L}}

\def \adss {$AdS_5 \times S^5$\ }

\def \D {\Delta}
\def \thet {\theta}
 \def \t {\tau}
 \def \p {\phi}
 \def \r {\rho}
 \def \rN {{\rm N}}
 \def\tw{{\tilde w}}
 \def\hJ{{J}}
 \def\hw{{w}}
 \def\hl{{\lambda}}
 \def\hth{{\theta}}
 \def\NN{{\cal N}}
 \def \bv {{ \bar w}}
\def \vn {{\vec n}}
\newcommand{\sfrac}[2]{{\textstyle\frac{#1}{#2}}}
\def \bl {{ \bar \lambda}}
\def \bp {{\bar p}}
\def \bu {{\bar u}}
\def \sha {\sfrac{1}{2}}
\def \w {\omega}
\def \ov {\over}
\def \vl { \vec \ell}
\def \varpi {{\rm w}}
\def \OO {{\cal O}}
\def \bG {\bar \G}

\def \c {\gamma}

\def \ss {{\rm s}}

\def \ve {\varepsilon}
\def \pa{\partial}
\def \I {{\cal I}}
\def \LL {{\cal L}}
\def \ep {\epsilon}
\def \R {{\rm R}}
\def \tilt {{\tilde t}}

\def\pic #1#2{\hbox{\lower#1pt\hbox{~\mbox{\epsfxsize=20truemm \epsffile{#2}}}}}

\def\pic #1#2#3{\hbox{\lower#1pt\hbox{~\mbox{\includegraphics[scale=#3]{#2}}}}}

\def \bt {\bar\theta}
\def \te {\theta}

\def \cc {{\rm f}}
\def \d {\delta}

\def \cL {{\cal L}}
\def \S  {{\rm S}}
\def \pp {{q}}
\def \vt {\vartheta}
\def \mm {{\cal  \ell}}
\def \Z {{\cal Z}}
\def \pa {\partial}
\def \C {{\cal C}}
\def \be {\bea}
\def \ee {\eea}
\def \c {\gamma}  \def \d {\delta}

\begin{document}
\overfullrule=0pt
\parskip=2pt
\parindent=12pt
\headheight=0in \headsep=0in \topmargin=0in \oddsidemargin=0in

\begin{center}
\vspace{0.1cm}
{\Large\bf
Matching the circular Wilson loop with \\
\vspace{0.15cm}
dual open string solution at $1$-loop in strong coupling\\
\vspace{0.2cm}

\vspace{0.3cm}
   }

 \vspace{.2cm} M. Kruczenski$^{}$\footnote{markru@purdue.edu}
  and A. Tirziu$^{}$\footnote{atirziu@purdue.edu}
\\
 \vskip 0.03cm

{\em
$^{}$Department of Physics, Purdue University,\\
525 Northwestern Ave., W. Lafayette, IN 47907-2036, USA\\
}

\end{center}

\begin{abstract}
We compute the $1$-loop correction to the effective action for the string solution in $AdS_5 \times S^5$ dual to the circular Wilson loop.
More generically, the method we use can be applied whenever the two dimensional spectral problem factorizes, to regularize
and define the fluctuation determinants in terms of solutions of one-dimensional differential equations. A such it can be applied to
non-homogeneous solutions both for open and closed strings and to various boundary conditions.
In the case of the circular Wilson loop, we obtain, for the $1$-loop partition function a result which up to a factor of two matches
the expectation from the exact gauge theory computation. The discrepancy can be attributed to an overall constant in the string partition
function coming from the normalization of zero modes, which we have not fixed.
\end{abstract}
\newpage

\renewcommand{\theequation}{1.\arabic{equation}}
 \setcounter{equation}{0}

\setcounter{equation}{0} \setcounter{footnote}{0}
\setcounter{section}{0}

\section{Introduction}

In recent years quite a number of tests were performed to check the AdS/CFT correspondence. Most of these tests involved matching
the anomalous dimension of certain operators in the gauge theory to the energy of a corresponding closed string in $AdS_5 \times S^5$.
Other possible way to check the correspondence is to compare the expectation values of Wilson loops in the gauge theory to the minimal area
of a string solution ending in the loop. In \cite{esz} it was conjectured that the expectation value of the circular Wilson loop in planar
$\mathcal{N}=4$ SUSY gauge theory can be obtained exactly to all orders in perturbation theory by summing all rainbow diagrams. Also, it
has been observed that the computation may be expressed in terms of a Gaussian matrix model. This conjecture was checked at one loop in
the gauge theory perturbative expansion in \cite{esz} and at two loops in \cite{aps,ps}.

The Gaussian matrix model was generalized further in \cite{dg} to obtain the expectation value of the circular Wilson loop to all orders
in the $1/N$ expansion and all orders in $\lambda=g^2 N$. Here $g^2$ is the gauge theory coupling parameter which is related to the string
coupling by $4 \pi g_s=g^2$. Very recently, it was shown in \cite{p} that the direct gauge theory computation to all orders precisely matches
the matrix model computation performed in \cite{dg}. The gauge theory prediction at all orders in $1/N$ was tested \cite{dg} against the leading
result at strong coupling in $\sqrt{\lambda}$ from string theory, and perfect matching was found. In this paper we test the gauge theory
prediction for the circular Wilson loop in the planar approximation, at the next order in the large-$\lambda$ expansion by comparing it to
the one loop correction to the partition function for the corresponding string solution.

Let us recall the expectation value of the circular Wilson loop at $N=\infty$ and all orders in $\lambda$ as it was obtained in \cite{esz,dg}
\begin{eqnarray}
<W>=\frac{2}{\sqrt{\lambda}}I_1 (\sqrt{\lambda})  \label{gc1}
\end{eqnarray}
where $I_1$ is a Bessel function. Expanding this for large $\lambda$ we obtain
\begin{eqnarray}
\ln <W>=\sqrt{\lambda} -\frac{3}{4}\ln \lambda + \frac{1}{2}\ln \frac{2}{\pi}-\frac{3}{8} \frac{1}{\sqrt {\lambda}}+  ... \label{gc}
\end{eqnarray}
The AdS/CFT correspondence relates the expectation value of circular Wilson loop to the string partition function for the corresponding string solution, i.e. $<W>=Z$. The logarithm of the string partition function can be written as
\begin{equation}
\ln Z = - \Gamma_0 - \Gamma_1+ ...
\end{equation}
where $\Gamma_0$ is the classical effective action, which is proportional to the area of the world-sheet, $\Gamma_1$ is the $1$-loop correction to the
effective action, and so on. Here we compute the string $1$-loop correction to the action by computing the string $2d$ effective action, $\Gamma_1= - \ln Z$. We want then to compare the string result to the expected expression from the gauge theory Wilson loop (\ref{gc}).

It was shown in \cite{dg} that the $\ln \lambda$ factor comes from the normalization of the zero modes in the string partition function. The numerical factor $\frac{1}{2}\ln \frac{2}{\pi}$ should come from the correct overall measure factor in the string partition function, and the contribution from the fluctuations of the sigma model near the corresponding string solution.

To determine the correct overall constant factor in the measure coming from the normalization of ghost zero modes\footnote{No zero modes arise from bosonic and fermionic fluctuations.} is difficult and it could be that can only
be done by comparison with the gauge theory. However, it should be noted that it depends only on the topology of the world-sheet and, in particular,
it is independent of the shape of the Wilson loop. On the other hand, the dependence on the shape is in the contribution from the $2d$ sigma model
fluctuation determinants which we compute here. It would be interesting to consider other closed loops since the ratio between their expectation
values should be independent of the zero mode normalization. In particular one can consider the string solutions corresponding to the circular $1/4$ BPS loops that were constructed in \cite{drukker}.

In the previously studied $1$-loop fluctuation determinants for the open string solution dual to the cusp Wilson loop \cite{krtt} or closed string solutions \cite{ft1,ft2,ptt,ftt}, the fluctuation Lagrangian had constant coefficients and one was able to find the spectrum of fluctuations easily. It turns out that the fluctuation Lagrangian near the string solution dual to the circular Wilson loop has non-constant coefficients. As we see below, this is the case even near the simple straight string solution. Finding the spectra of the corresponding operators becomes a nontrivial task. Both the straight string and circular string were discussed in \cite{dgt}. The extension to the parallel lines solution was discussed in \cite{dgt,fgt}, and a supersymmetric extension of the circular Wilson loop was discussed in \cite{z}. Fluctuations near these open string solutions were discussed in \cite{dgt,fgt,z}. However, there were no attempts to explicitly compute the functional determinants and therefore obtain explicit results for the $1$-loop partition functions for these solutions.

It is the purpose of this paper to compute the $1$-loop effective action for the straight and circular Wilson loop string solutions. One motivation is the computation in itself of the $1$-loop fluctuation determinants for such cases where the fluctuation Lagrangian has non-constant coefficients. Another motivation in the particular case of the circular Wilson loop is the comparison to the gauge theory expectation in (\ref{gc}).

The open string computations in these cases present extra challenges due to the presence of the linear divergency expected at the boundary.
It was shown in \cite{dgo} that at the classical level the linearly divergent term is proportional to the length of the Wilson loop.
To get rid of the IR linear divergent\footnote{From the field theory point of view this is a UV divergence.} term at the classical level, one may just
subtract all terms proportional to $\frac{1}{\epsilon}$, where $\epsilon$ is a IR worldsheet cutoff.
This prescription could be extended also at the $1$-loop level. However, a better and less ambiguous way to treat the $\epsilon\rightarrow 0$
divergency is to subtract a reference solution so that the divergency cancels. As we see in this paper, this is a consistent method at the classical level and $1$-loop. Although we do not go beyond $1$-loop here, it is likely that this method of subtracting a reference solution is consistent also
at higher loops (at strong coupling). In this paper the reference solution is the straight string, whose value is subtracted from the circular
Wilson loop producing a finite result.

For the computation of the ratios of determinants that appear in the $1$-loop correction we employ a method put forward long ago in \cite{gy}, which was developed and improved recently in a series of papers \cite{kl,km1,mt}. The method was used before in field theory computations \cite{dk,dhlm,dm}; for a recent review of those computations see also \cite{d}. In short the method says that the ratio of two one-dimensional determinants is the ratio of the respective (non-normalizable) wave-functions corresponding to the zero eigenvalue of the operators and evaluated at the boundary. We present a review of this method in Appendix A. In order to use this method the two-dimensional spectral problems involved must be separable into one-dimensional ones. Then the only remaining problem for an arbitrary such string
solution is to find the solutions of the relevant one dimensional differential equations, which even when they cannot be solved exactly,
might be computed using numerical methods. An interesting further application of this method is in the case of the two parallel lines \cite{dgt}.

The Wilson loop can be extended to have a winding number $k$ along the circle. It was conjectured in \cite{df}, and directly proved in \cite{p} that the gauge theory partition function is
\begin{eqnarray}
<W>_k=\frac{2}{k \sqrt{\lambda}}I_1 (k \sqrt{\lambda})
\end{eqnarray}
This is the same as the expression for winding $k=1$, (\ref{gc1}) with the replacement $\lambda\rightarrow k^2 \lambda$. The expansion of the logarithm of this partition function at large $\lambda$ is
\begin{eqnarray}
\ln <W>_k=k \sqrt{\lambda} -\frac{3}{4}\ln \lambda - \frac{3}{2}\ln k + \frac{1}{2}\ln \frac{2}{\pi}-\frac{3}{8 k} \frac{1}{\sqrt {\lambda}}+  ...
\label{gck}
\end{eqnarray}
One would like to compare (\ref{gck}) to the corresponding dual string solution. However, a technical complication appears in the case $k \neq 1$.
The dual string solution in this case has a worldsheet surface whose boundary winds around the circle $k$-times. Such surface wraps $k$ times
the surface corresponding to $k=1$ but topologically is still a disk with Euler number equal to $1$. As a result there is a conical singularity
at the center of the disk.

We extend the $k=1$ computation of the $1$-loop effective action to the case with arbitrary $k$. In our method of computing the determinants we
effectively cut the origin of the disk\footnote{We put Dirichlet boundary conditions at the new boundary.}, compute the determinants and then analytically extend the computation to the whole disk including the
origin. While this procedure is not problematic for $k=1$, for arbitrary $k$ it might miss a relevant contribution from the conical singularity.
With this caveat, we can compare our result with (\ref{gck}). We will see that it does not agree so the understanding of the
$k \neq 1$ case remains an interesting problem which needs to be clarified.

Let us mention that a further very useful comparison between the gauge theory prediction at strong coupling (\ref{gc}) and string string theory is at two loops in strong coupling. In that case one does not need to know the precise numerical coefficient from the zero modes. However, while such two loop computations were performed recently \cite{rtt,rt} for homogenous solutions, it seems hard to do such a computation in the non-homogenous cases such as the Wilson circular loop solution. Nevertheless, such a $2$-loop computation would be very interesting to perform as it would be a further very useful check of the AdS/CFT correspondence.

We start below by reviewing the straight  string solution and computing the $1$-loop correction to the effective action using the ratio of determinants method. We then compute the $1$-loop effective action for the circular Wilson loop solution. We generalize this computation also to an arbitrary winding number $k$. In appendix A we present in some detail the method of computing ratios of determinants employed in this paper, and a simple application of the method to free massive fields.

\renewcommand{\theequation}{2.\arabic{equation}}
 \setcounter{equation}{0}

\section{Straight string solution}

The $1$-loop correction to the partition function of the straight string solution in $AdS_5 \times S^5$ was considered in  \cite{dgt}. In what follows we work in conformal gauge with $\sqrt{g}g^{ij}=\delta^{ij}$ and use the Polyakov action.
 The straight string solution is
\begin{equation}
x_0=\tau, \quad \quad z=\sigma  \label{st}
\end{equation}
with the $AdS_5$ metric being
\begin{equation}
ds^2=\frac{1}{z^2}(d x_0^2 +d x_1^2 +dx_2^2 +dx_3^2+dz^2)  \label{mcc}
\end{equation}
As the radial coordinate $z$ runs from the boundary of $AdS_5$ to its horizon, the worldsheet coordinate $\sigma$ takes values in the interval $0 \leq \sigma < \infty $. We periodically identify $x_0$ in the interval $0\leq x_0 \leq 2 \pi T$, where $T$ is taken to be large (i.e. $0\leq \tau \leq 2 \pi T$). The minimal surface for this string solution is a half plane extended along the $x_0$ line and $z \geq 0$.

The induced metric on this solution is that of $AdS_2$
\begin{equation}
ds^2_2=\frac{1}{\sigma^2}(d \tau^2 + d \sigma^2)
\end{equation}
while the corresponding $2d$ curvature $R^{(2)}=-2$. The classical value of the action corresponding to this string solution is
\begin{equation}
S=\sqrt{\lambda}\frac{T}{\epsilon}  \label{tst}
\end{equation}
where, as in \cite{dgo}, we introduced a cutoff, $\epsilon$, near the boundary of AdS.

Let us observe that the action obtained in (\ref{tst}) is just proportional to the volume part of the Euler characteristic, i.e
\begin{equation}
S= -\sqrt{\lambda}\chi_v, \quad \quad
\chi_v=\frac{1}{4 \pi}\int_M d^2 \sigma \sqrt{g}R^{(2)}= - \frac{T}{\epsilon}, \quad \quad \chi_b=\frac{1}{2 \pi}\int_{\partial M}d s \kappa_g= \frac{T}{\epsilon}
\end{equation}
where we denoted by $\chi_v$ and $\chi_b$ the volume and boundary parts of the Euler number, while $M$ is a general $2d$ surface. This is not a specific feature of the straight string but it rather happens also for the circular string and other situations in general that have constant $2d$ curvature. At $1$-loop level, the UV divergency also turns out to be proportional to the volume part of the Euler number.

At the classical level one way to get rid of the linear divergency is to consider a Legendre transform  \cite{dgo,df1}. In terms of Euler characteristic this amounts to include its boundary term, so that the finite action is just \cite{dgt,dg}
\begin{equation}
S= -\sqrt{\lambda}\chi, \quad \quad
\chi = \chi_v + \chi_b  \label{iis}
\end{equation}
This action is zero for the half-plane, because $\chi=0$ for a half plane. As it should, the boundary term in the Euler number cancels the volume term.
More generically for any smooth loop, the boundary term in the Euler number will always be singular and proportional to the length of the Wilson
loop $\frac{L}{\epsilon}$ ($L$ is the length of the loop). The completion of the Euler number appears to provide a natural way to regularize the
classical result\footnote{At $1$-loop, however, this simple procedure seems not to be enough, which is why we choose to regularize by subtracting a reference solution.}.

For completness, let us mention that in \cite{dgo,df1} another method of regularizing the area was proposed. Essentially, it consists in taking the Legendre transform of the action. More precisely, one adds to the Lagrangian a total derivative,
\begin{equation}
\tilde{L}=L +\partial_{\sigma}[z \frac{\partial L}{\partial(\partial_{\sigma} z)}]
\end{equation}
such that the new action becomes
\begin{equation}
\tilde{S}= S  - \epsilon \int d \tau \frac{\partial L}{\partial(\partial_{\sigma} z)}\bigg|_{z=\epsilon} \label{bcd}
\end{equation}
where $S$ is the original Polyakov action and $\tau$ denotes the boundary coordinate. Throughout this paper we consider the boundary to be at some small but finite cutoff $z=\epsilon$. In the case of the straight string using $\frac{\partial L}{\partial(\partial_{\sigma}z)}= \frac{\sqrt{\lambda}}{2 \pi}\frac{\partial_{\sigma}z}{z^2}$, the transformed action is
\begin{equation}
\tilde{S}=S - \sqrt{\lambda}\frac{T}{\epsilon}
\end{equation}
Indeed the last term cancels the linearly divergent term in (\ref{tst}) making the action finite.

\renewcommand{\theequation}{3.\arabic{equation}}
 \setcounter{equation}{0}

\section{Straight string: one loop correction to the effective action}

We want to compute the one loop correction to the effective action for the straight string. This was done in \cite{dgt} by an indirect method using the results of \cite{st}. However, divergencies were present and $\zeta$ function regularization was used to show that the $1$-loop correction to the effective action is zero. Here we do this computation by directly dealing with the determinants that appear in the partition function. To do this, one needs the spectra of quadratic bosonic and fermionic fluctuations near the solution. The general expressions of the bosonic and fermionic fluctuation operators have been computed in \cite{dgt}. Let us start by reviewing the bosonic fluctuation action near any classical solution. Introducing the tangent-space components of fluctuations as
\begin{equation}
x^{\mu}\rightarrow \bar{x}^{\mu}+ \xi^{\mu}, \quad \quad \zeta^{A}=E^{A}_{\mu}\xi^{\mu}, \quad A=0,1,...9
\end{equation}
as well as fluctuations near a background metric
\begin{equation}
g_{ij}\rightarrow g_{ij}+ \chi_{ij}
\end{equation}
one obtains the following quadratic action in conformal gauge \cite{dgt}
\begin{eqnarray}
S=\frac{\sqrt{\lambda}}{4 \pi}\int d^2 \sigma \sqrt{g}\bigg[g^{ij} D_i \zeta^a D_j \zeta^a + X_{ab} \zeta^a \zeta^b + g^{ij} D_i \zeta^p D_j \zeta^p+X_{pq} \zeta^p \zeta^q\bigg]\\ \label{lagb}
X_{ab}= -g^{ij} e^c_i e^d_j R_{abcd}, \quad X_{pq}=-g^{ij} e^r_i e^s_j R_{pqrs}\nonumber
\end{eqnarray}
where the indices $a,b=0,1,2,3,4$ refer to the $AdS_5$ space while $p,q=5,6,7,8,9$ to the sphere $S^5$. Also, here
\begin{equation}
e^a_i= E^a_{\mu} \partial_i \bar{x}^{\mu}, \quad e^p_i=E^p_{\mu} \partial_i \bar{x}^{\mu}
\end{equation}
are the projections of the $AdS_5$ and $S^5$ vielbeins on the worldsheet. $D_i$ is the covariant derivative containing the projection of the target space spin connection on the worldsheet
\begin{equation}
D_i \zeta^a =\partial_i \zeta^a+ \partial_i \bar{x}^{\mu} \Omega^{ab}_{\mu} \zeta^b
\end{equation}
The fluctuation fields have canonical norms
\begin{equation}
\|\zeta^a\|^2=\int d^2 \sigma \sqrt{g}\zeta^a \zeta^a, \quad \quad  \|\zeta^p\|^2=\int d^2 \sigma \sqrt{g}\zeta^p \zeta^p
\end{equation}
For the metric (\ref{mcc}) the projections of vielbeins and spin connection are
\begin{eqnarray}
e^a_i=\bigg(\frac{1}{z}\partial_i x_0, \frac{1}{z}\partial_i x_1, \frac{1}{z} \partial_i x_2, \frac{1}{z}\partial_i x_3, \frac{1}{z}\partial_i z\bigg)\nonumber\\
\Omega^{\alpha}_{i 4}=-\frac{1}{z}\partial_i x_{\alpha}, \quad \alpha=0,1,2,3  \label{vsps}
\end{eqnarray}
The bosonic ghost action for the $2d$ vectors is \cite{dgt}
\begin{equation}
 S_{gh}=\frac{\sqrt{\lambda}}{4 \pi}\int d^2 \sigma \sqrt{g}g^{ij}\bigg(g^{kl} \nabla_k \epsilon_i \nabla_l \epsilon_j - \frac{1}{2}R^{(2)} \epsilon_i \epsilon_j \bigg)  \label{ghosts}
 \end{equation}
where $\nabla$ includes the worldsheet connection.

In general, due to the Weyl symmetry, the classical value of the metric can be taken to differ from the induced metric, $h_{ij}$, by an arbitrary conformal factor $\rho$, $g_{ij}=e^{2 \rho} h_{ij}$. The bosonic fluctuation Lagrangian (\ref{lagb})  was obtained assuming an arbitrary  background metric $g_{ij}$,
but, in order to proceed with the computation, we need to choose a specific metric $g_{ij}$. Two obvious possibilities are the induced metric and the
flat metric. Certainly, the physical finite part of the result should not depend on the background metric used. Let us now consider the benefits of each choice.

 As argued in \cite{dgt}, in the Green-Schwarz (GS) formulation, at $1$-loop in $AdS_5 \times S^5$, the logarithmic divergencies can be shown to cancel by
the same argument used in flat space GS action. The overall factor from the measure in the partition function has a logarithmic UV divergency that depends on the Euler characteristic
\cite{l,a, dgt}, $ e^{- 3 \chi \ln \Lambda}$, (where $\Lambda$ is a dimensionless large cutoff), which comes from the cutoff dependent factors in the conformal Killing vectors and/or Teichmuller deformations, as a result of fixing the world-sheet reparametrization symmetry and the Weyl symmetry\footnote{The $\kappa$ symmetry ghosts after fixing the $\kappa$ symmetry does not give a logarithmic UV divergency.}. As it was shown in the general case in \cite{dgt}, this divergency is canceled by the logarithmic divergence of the one loop determinants.
How precisely, this UV divergency is canceled by the contribution from the $1$-loop fluctuation determinants is rather subtle.

There are two sources of UV divergency from the $1$-loop determinants: one that is proportional to the curvature $R^{(2)}$ of the background metric $g_{ij}$, and the other one comes from the background field. In the case when $g_{ij}$ is the induced metric the cancelation of UV divergency is rather obscure as the divergencies coming from the two sources mix. The overall result from the one loop determinants may not be zero, and it is canceled by the one from the measure. To obtain this cancelation one needs to consider another subtle fact pointed out in \cite{dgt}. In GS formulation the quadratic fermionic Lagrangian can be brought after rotations and field redefinitions in a form similar to eight $2d$ fermions. However, the norm of this actually GS fermions is different from true $2d$ fermions. To change the norm of a $2d$ fermion into the norm of a GS fermion one needs a local determinant, which brings a factor of four into the curvature, $\int R^{(2)}$, part of the UV divergency. In other words a GS fermion contributes four times more to the topological ($\int R^{(2)}$) UV divergency than a $2d$-fermion. This appears already in the GS string in flat space. In this paper it is convenient for the practical purpose to take the fermionic determinant as that of $2d$ determinants. For the solely purpose of practical computation of the finite part of the $1$-loop correction, we will not be concerned about this local determinant, as UV divergency cancelation was shown already in general in \cite{dgt}.

In the case when $g_{ij}$ is the induced metric, the UV divergency,  $\int R^{(2)}$, may not be zero, and so may be the divergency coming from the background fields. Even though in this case the result from the $1$-loop fluctuation determinants is not UV finite by itself, this choice of $g_{ij}$ being the induced metric, as in \cite{dgt,fgt}, appears to be the best choice for the case of open strings with boundaries, and also for our method of computing functional determinants. Thus, we will fix the background metric $g_{ij}$ to be the induced metric throughout this paper.

When one chooses $g_{ij}$ flat, the UV divergency coming from $\int R^{(2)}$  is zero, while the one coming from the background field cancels between bosons and fermions \cite{dgt}. In this case, assuming also that one can ignore the boundary terms, the contribution from the $1$-loop determinants to the UV divergency should vanish by itself, so one expects that the contribution from the one loop fluctuation determinants will be UV finite. This is indeed the case as we have checked using the computation of ratio of determinants employed in this paper. It turns out that within our method of computing the functional determinants, the choice of flat $g_{ij}$ is more cumbersome since it leads to complicated longitudinal mode fluctuation operators and issues with the appropriate boundary conditions for them. Complications from the proper choice of boundary conditions seems to appear in the case of circular string.

\bigskip

\bigskip

After the general discussion about the cancelation of UV divergency let us return to the case of the straight string. As we pointed out we take $g_{ij}$ to be the induced metric. The only non-trivial covariant derivatives are
\begin{equation}
D_0 \zeta^0 =\partial_0 \zeta^0 - \frac{1}{\sigma}\zeta^4, \quad \quad D_0 \zeta^4=\partial_0 \zeta^4 + \frac{1}{\sigma}\zeta^0
\end{equation}
 while the mass matrix is $X_{ab}= \mbox{diag}(1,2,2,2,1)$ and $X_{pq}=0$. One can show \cite{dgt} that the ghost action is identical to the action of the longitudinal modes $\zeta^0,\zeta^4$ so, in the partition function their contributions cancel each other. The remaining transverse bosonic fluctuations have all masses squared equal to $2$. From the sphere fluctuations one obtains
$5$ massless modes. The quadratic transverse fluctuation action is
\begin{equation}
S=\frac{\sqrt{\lambda}}{4 \pi}\int d \tau d \sigma \frac{1}{\sigma^2}\bigg[\sigma^2\partial_i \zeta^A \partial_j \zeta^A +2 (\zeta^1)^2+
2(\zeta^2)^2+2(\zeta^3)^2\bigg]
\end{equation}
where here $A=1,2,3,5,6,7,8,9$, i.e. the longitudinal bosonic fluctuations corresponding to directions $\zeta^0,\zeta^4$ are excluded.
Therefore, the resulting spectral problem one needs to solve is
\begin{equation}
L f = \Lambda f, \quad L=\sigma^2(-\partial_0^2-\partial_1^2)+2 \label{stf2}
\end{equation}
and the same problem but with mass equal to zero for the sphere fluctuations.
Since we want to compare the results between straight and circular string solutions we choose periodic boundary condition in $\tau$. In $\sigma$ we choose Dirichlet boundary conditions. Using the expansion $f(\tau,\sigma)=\sum_{n}g_n (\sigma) e^{i m \tau} $ with $m=\frac{n}{ T}$ (recall that $0 \leq \tau < 2 \pi T$) where $n$ is an integer number, we obtain the spectral problem to be solved for each $m$
\begin{equation}
\sigma^2(- g''+ m^2 g)+2 g =\Lambda g
\end{equation}
To obtain the determinant of $L$ one needs to take a product over $m$
\begin{equation}
\det L= \prod_{m} \det \bigg(\sigma^2 (-\partial_1^2 +m^2) + 2\bigg)
\end{equation}
Since $T$ is taken to be large, as appropriate for the straight string, we replace at the end the sum over $m$ by an integral which gives
the dominant term in the $T\rightarrow \infty$ limit. Before considering solving this problem let us see first what operator comes out from
the fermionic Lagrangian.

\bigskip

\bigskip

Let us now consider the fermionic contributions. Before fixing $\kappa$
symmetry the fermionic Lagrangian is
\begin{equation}
S_F= \frac{\sqrt{\lambda}}{2 \pi}\int d^2 \sigma L_{2F}, \quad \quad L_{2F}=-i (\sqrt{g}g^{ij} \delta^{IJ}-\epsilon^{ij}s^{IJ}) \bar{\theta}^I \rho_i D_j \theta^J
\end{equation}
where the spinors $\theta_1$ and $\theta_2$ are $16$-component real Majorana-Weyl fermions of the same chirality, and
\begin{equation}
\rho_i=\Gamma_A e_i^A
\end{equation}
\begin{equation}
D_i \theta^I=\delta^{IJ} \nabla_i-\frac{1}{2}i \epsilon^{IJ}\rho_i \theta^J, \quad \nabla_i=\partial_i+\frac{1}{4}\Omega_i^{AB}\Gamma_{AB}
\end{equation}
Following \cite{dgt} we fix $\kappa$ symmetry by taking $\theta_1=\theta_2$ and the fermionic Lagrangian becomes
\begin{equation}
L_{2F}=-2 i \sqrt{g}g^{ij} \bar{\theta}\rho_i \nabla_j \theta+ \epsilon^{ij} \bar{\theta}\rho_i \rho_j \theta  \label{mnm}
\end{equation}
Let us mention that upon fixing $\kappa$ symmetry, $\kappa$ symmetry ghosts arise. However, their contribution does not give a logarithmic divergency; only power divergent terms and possibly a finite part appear. In dimensional regularization the net contribution from the $\kappa$ symmetry ghosts is zero. With a cutoff regularization, power divergencies should cancel those coming from the conformal Killing vectors/Teichmuller spaces, while any remaining finite part contributes to the overall numerical coefficient in the string partition function, which we do not fix in this paper. 

The quadratic GS fermionic action (\ref{mnm}) has exactly the same form as the action for $2d$ fermions in curved $2d$ space. As in \cite{dgt} here we use Minkowski metric, and at the end we will switch back by taking $\partial_0\rightarrow i \partial_0$. Also, we take the background metric to be the induced metric like in the bosonic case.

In the case of the straight string one obtains \cite{dgt}
\begin{equation}
\nabla_0 = \partial_0 - \frac{1}{2\sigma}\Gamma_{04}, \quad \quad \nabla_1 = \partial_1, \quad \rho_0=\frac{1}{\sigma}\Gamma_0, \quad \rho_1 =\frac{1}{\sigma}\Gamma_4
\end{equation}

The fermionic Lagrangian is
\begin{equation}
L_{2F}=- 2 i \sqrt{g} \bar{\theta} D_{F} \theta
\end{equation}
where
\begin{equation}
D_{F}=-\sigma \Gamma_0 \partial_0 + \sigma \Gamma_4 \partial_1 -\frac{1}{2}\Gamma_4+i \Gamma_0 \Gamma_4
\end{equation}
We assume the standard normalization of fermions $\|\theta\|^2 =\int d^2 \sigma \sqrt{g}\bar{\theta}\theta$. The matrices $\Gamma_0, \Gamma_4$ play the role of worldsheet $2d$ Dirac matrices, since we can choose the following
 representation $\Gamma_0=i \sigma_2 \times I_8, \Gamma_4=\sigma_1 \times I_8$, where $\sigma_{1,2}$ are Pauli matrices.
Squaring the above fermionic operator or computing its determinant using the above gamma matrix representation, one obtains the
spectral problem to be solved. One then ends up with the following spectral problem for the fermions
\begin{equation}
L_{F} \theta = \Lambda \theta
\end{equation}
where the $2 \times 2$ operator is
\begin{equation}
L_{F}=-\nabla_i \nabla^i + \frac{R^{(2)}}{4}+1=\sigma^2 (-\partial_1^2+m^2)+\frac{3}{4}+\Gamma_{04} m \sigma \label{stf}
\end{equation}
As in the bosonic sector, we again took the $\tau$ part of the solution of the form $e^{i m \tau}$, with $m=\frac{n}{T}$.
Since $\Gamma_{04}$ is diagonal with elements $1$ and $-1$, the determinant of $L_F$ is a product of two one dimensional determinants.

\bigskip

Putting together the bosons and fermions one ends up with the following $1$-loop partition function \cite{dgt}\footnote{This expression for
the partition function can also be obtained by starting with the Nambu action and fixing the fluctuations of the two longitudinal fields to
zero \cite{dgt}. There is no ghost Lagrangian in the Nambu method, and the overall measure factor is computed differently than in conformal
gauge. In any case, in this paper we only compute the contribution from the fluctuation determinants.}
\begin{equation}
Z=\frac{\det^{8/2} (-\nabla^2 + \frac{R^{(2)}}{4}+1)}{\det^{3/2}(-\nabla^2 +2) \det^{5/2}(-\nabla^2)} \label{part}
\end{equation}
where Laplace operators in curved space are $\nabla^2= \frac{1}{\sqrt{g}}\nabla_i (\sqrt{g}g^{ij}\nabla_j)$. More precisely, taking $g$ the induced metric and $R^{(2)}=-2$  we obtain
\begin{eqnarray}
Z= \prod_{m} \ \frac{\det^{4/2}[\sigma^2(-\partial_1^2 + m^2)+ \frac{3}{4}+ m \sigma] \ \det^{4/2}[\sigma^2(-\partial_1^2 + m^2)+ \frac{3}{4}- m \sigma] }{\det^{3/2}[\sigma^2 (-\partial_1^2 + m^2)+2] \ \det^{5/2}[\sigma^2 (-\partial_1^2 + m^2)]} \label{oes}
\end{eqnarray}

\bigskip

The computation of each of these determinants is difficult in general as they are infinite and one has to deal  with divergencies. Here we employ a method of computing ratio of determinants which gives a finite result for each particular ratio. We review this method in the Appendix A. Let us just summarize here the method.
The ratio of the determinants of two second-order differential operators $M_1,M_2$ defined on the interval $[0,\infty)$, and
satisfying Dirichlet boundary conditions can be computed as follows \cite{gy,c,kl}
\begin{equation}
\frac{\det(M_1)}{\det(M_2)}= \lim_{R\rightarrow \infty} \frac{\psi_1 (R)}{\psi_2 (R)} \label{ratio}
\end{equation}
where $\psi_i$ satisfy the initial value problems
\begin{equation}
M_i \psi_i = 0, \quad \quad \psi(0)=0, \quad \psi'(0)=1 \label{inv}
\end{equation}
The operators $M_1,M_2$ are of the form
\begin{equation}
M_i= -\frac{d^2}{d x^2}+V_i(x), \quad \quad i=1,2
\end{equation}

We want to apply this method to compute the ratio of determinants as they appear in (\ref{part}). As shown in Appendix A (\ref{yef}), the ratio of determinants with rescaled operators is the same as the ratio with the coefficient of second order derivative not rescaled to $1$  because the initial value problems (\ref{inv}) are the same. From now on for convenience when we compute the initial value solutions we always consider the operators with $g^{00}$ scaled away; again ratio of determinants are the same with $g^{00}$ scaled away or not. It was shown in \cite{ch} that in curved spaces the $1$-loop correction to the vacuum energy can be computed from the determinants that have the factor $g^{00}$ in front of $\partial_0^2$ scaled away.  Thus in our case we conclude that the $1$-loop correction to the vacuum energy is like in flat space, i.e. given by $E_1 \sim -\ln Z$. This was also shown in \cite{dgt} by directly using the expansion of determinants defined through heat kernel.

An important fact in computing the ratio of determinants that appears in (\ref{oes}) is the presence of a singularity at $\sigma=0$. The initial value problem contains a singularity at the boundary $\sigma=0$. This is a complication as we would like to get a finite result in the physical interval $[0,R]$. However, we expect such a complication to arise since it is already present at the classical level. At the classical level the prescription was to add a boundary term, or, equivalently, to complete the Euler number. It is not clear what mechanism one is to use at the $1$-loop level to subtract this divergence. Below we will take the interval for $\sigma$ to be $[\epsilon,R$], and in order to get rid of the $\frac{1}{\epsilon}$ divergency we subtract the results of two different solutions, i.e. straight and circular string solutions.

Before computing the determinants in (\ref{oes}) with the method reviewed above, let us mention that we are computing the GS fermionic determinants in (\ref{oes}) as if they were $2d$ fermions and not GS fermions. As discussed already, the difference between the two come only from the different norm, which translates into a local determinants that does not affect the finite part of the result.

Let us write the $1$-loop effective action as
\begin{equation}
\Gamma_1= \frac{1}{2}\ln \prod_m P_m
\end{equation}
where $m=\frac{n}{T}$, $n$ being an integer number, and
\begin{eqnarray}
P_m= \frac{\bigg[\det[-\partial_1^2+m^2 +\frac{2}{\sigma^2}]\bigg]^3 \bigg[\det[-\partial_1^2+m^2]\bigg]^5}{\bigg[\det[-\partial_1^2+m^2 +\frac{3}{4 \sigma^2}+ \frac{m}{\sigma}\bigg]^4 \bigg[\det[-\partial_1^2+m^2 +\frac{3}{4 \sigma^2}- \frac{m}{\sigma}\bigg]^4}
\end{eqnarray}
In view of the symmetry $m\rightarrow -m$, we restrict to the $m \geq 0$ case. We keep $m$ in all formulas, and only at the end replace it
in terms of $n$. In fact, since $T$ is large one can replace the sum over $n$ by an integral.

The next step is to compute the ratio of the above determinants using (\ref{ratio}). As we already pointed out, we take Dirichlet boundary conditions
in $\sigma$ and compute the ratios of determinants in the interval $\sigma\in[\epsilon,R]$ with $R$ large. At the end, we take the $R\rightarrow\infty$
limit. The introduction of a large but finite $R$ effectively introduces another boundary for the worldsheet. This boundary, however, is un-physical
and, in fact, all $R$ dependence goes away at the end. Besides, the boundary part of the Euler number corresponding to this additional boundary vanishes
when $R\rightarrow \infty$, thus, this extra boundary has no effect on the Euler number.

As in the classical area we expect a linear divergency near the boundary of $AdS$; so, to account for this we again considered a small cutoff $\epsilon$. Only at the very end in the expression of the $1$-loop energy we take the limit $\epsilon \rightarrow 0$. We need to solve the following initial value problems
\begin{equation}
-g''+\bigg(m^2+\frac{2}{\sigma^2}\bigg) g = 0
\end{equation}
with the initial conditions
\begin{equation}
g(\epsilon)=0, \quad g'(\epsilon)=1
\end{equation}
The solution is
\begin{equation}
g(\sigma)=\frac{1}{m^3 \epsilon \sigma}\bigg[m(\sigma-\epsilon)\cosh m(\sigma-\epsilon)-(1- \epsilon m^2 \sigma)\sinh m(\sigma-\epsilon)\bigg]
\end{equation}
We also need to solve
\begin{equation}
\bigg[ -\partial_1^2+m^2+\frac{3}{4 \sigma^2}+\frac{m}{\sigma}\bigg]\theta=0
\end{equation}
whose solution is
\begin{eqnarray}
\theta(\sigma)=\frac{1}{4 m^2 \sqrt{\epsilon \sigma}}\bigg[(2 m \sigma -1) e^{m (\sigma-\epsilon)}- (2 m \epsilon-1) e^{-m (\sigma- \epsilon)}\bigg]
\end{eqnarray}
Lastly for the free bosons we need
\begin{equation}
-g''+m^2 g = 0
\end{equation}
with solution
\begin{equation}
g= \frac{1}{m}\sinh m(\sigma-\epsilon)
\end{equation}

Taking the large $\sigma$ limit of these solutions we obtain the following finite ratios of determinants
\begin{equation}
\frac{\det[-\partial_1^2+m^2 + \frac{2}{\sigma^2}]}{\det[-\partial_1^2 +m^2 +\frac{3}{4 \sigma^2}+\frac{m}{\sigma}]}=\frac{m \epsilon +1}{m \sqrt{\epsilon R}}, \quad
\frac{\det[-\partial_1^2+m^2]}{\det[-\partial_1^2 +m^2 +\frac{3}{4 \sigma^2}+\frac{m}{\sigma}]}=\sqrt{\frac{\epsilon}{R}}
\end{equation}
\begin{equation}
\frac{\det[-\partial_1^2+m^2]}{\det[-\partial_1^2 +m^2 +\frac{3}{4 \sigma^2}-\frac{m}{\sigma}]}=\frac{2 m \sqrt{R \epsilon}}{2m \epsilon+1}
\end{equation}
These formulas are valid in the large $R$ limit but we have been careful not to take $R>m$, as $m$ can be large in the sum.

Putting the ratios together we obtain
\begin{eqnarray}
\ln \prod_{m \neq 0}^{\infty}P_m= \sum_{m\neq 0} \ln \bigg[\frac{16 m \epsilon (1+ m \epsilon)^3}{(1+ 2 m \epsilon)^4}\bigg]
\end{eqnarray}
It is interesting and crucial that in the above final formula, the $R$ dependence cancels, therefore taking the
limit $R\rightarrow\infty$ gives a finite result. Note that for generic determinants this is not always true, here it is a reflection
of the fact that the boundary at $\sigma=R$ is unphysical.

As we already pointed out, with periodic boundary conditions on $\tau$, $m=\frac{n}{T}$, and in the large $T$ limit one should replace the sum over $n$ by an integral\footnote{The same result is obtained by doing the sum first and then keeping the leading terms in large $T$.}. Therefore the $1$-loop effective action is
\begin{equation}
\Gamma_1= \int_0^{\infty} d n \ln \bigg[\frac{ n (n+ \frac{T}{\epsilon})^3}{(n +\frac{T}{2 \epsilon})^4}\bigg] \label{ints}
\end{equation}
This integral is UV logarithmically divergent as we expect since the volume part of the Euler characteristic is not zero when $g_{ij}$ is the induced metric. Introducing a large dimensionless cutoff, $\Lambda$, in the summation indices we obtain.
\begin{equation}
\Gamma_1= \int_0^{\Lambda} d n \ln \frac {n (n+\frac{T}{\epsilon})^3}{(n +\frac{T}{2 \epsilon})^4}=\frac{T}{\epsilon}(1 + \ln \frac{\epsilon}{4 T}) + \frac{T}{\epsilon}\ln \Lambda \label{1loops}
\end{equation}

The UV divergent part of $\Gamma_1$ is proportional to the volume part of the Euler characteristic
\begin{equation}
\Gamma_1 \ \ \rightarrow \\  - \chi_v \ln \Lambda, \quad \quad \chi_v=\frac{1}{4 \pi} \int_0^{2 \pi T} d \tau \int_{\epsilon}^{\infty} d \sigma \sqrt{g}R^{(2)} = -\frac{T}{\epsilon}
\end{equation}
where in this computation $g_{ij}$ is the induced metric $ds^2=\frac{1}{\sigma^2}(d \tau^2 + d \sigma^2)$, and $R^{(2)}=-2$. As discussed already, we have used $2d$ fermions in the computation of the determinants so we do not expect precise UV divergency cancelation between this result and the one from the measure factors. Using the Seeley coefficients obtained from the expansion of the heat kernel expression of functional determinants, we checked that the
UV divergency we obtain using the ratio of determinant method indeed gives the same result\footnote{For a bosonic operator $-\nabla^2 +X$, the volume part of the Seeley coefficient is \cite{dgt}
\begin{equation}
b_2= -\frac{R^{(2)}}{6}+X
\end{equation}
while for a $2d$ Majorana fermion with squared Dirac operator $- \nabla^2 + Y$, it is
\begin{equation}
b_2= \frac{R^{(2)}}{12}+Y
\end{equation}
We have eight transverse bosons, with total mass $6$, and eight $2d$ fermions each with mass squared equal to $1$. The total Seeley coefficient
in the partition function is then ($R^{(2)}=-2$ in our cases)
\begin{equation}
 8 \frac{R^{(2)}}{12}+ 8 + 8 \frac{R^{(2)}}{6}-6 = 2 R^{(2)}+2 = -2
\end{equation}
The UV divergent part in the logarithm of the partition function is given by
 \begin{equation}
\frac{1}{2} \frac{1}{4 \pi}\ln \Lambda^2 \int d^2 \sigma  \sqrt{g} (-2)= \chi_v \ln \Lambda
\end{equation}
 The UV divergence of the 1-loop effective action is then $\Gamma^{\text{div}}_1 = - \chi_v \ln \Lambda $. This is the same as what we obtained by using
the ratio of determinants method.
For completeness, let us recall the counting when instead of a $2d$ fermions one has a GS fermion (divergency from $\int R^{(2)}$ has an additional factor of $4$)
\begin{equation}
4\times 8 \frac{R^{(2)}}{12}+ 8 + 8 \frac{R^{(2)}}{6}-6 = 4 R^{(2)}+2 = -6 = 3 R^{(2)}
\end{equation}
This gives a UV divergency $3 \chi_v \ln \Lambda$ in $\ln Z$, which (after adding a corresponding boundary term) indeed cancels the corresponding divergency from the measure. }. At the classical level we added a boundary term to complete the Euler number. Boundary terms are
always $\frac{1}{\epsilon}$ type terms. While we do not have a precise mechanism at $1$-loop to regularize the $\frac{1}{\epsilon}$ divergency
we again should add a term to complete the Euler number, so the UV divergency is $-\chi \ln \Lambda$.

 The final result (\ref{1loops}) is still divergent in the limit $\epsilon\rightarrow 0$. However, since we know that the straight string
is BPS, the correct result is zero, namely
\begin{equation}
Z=1  \label{1loopss}
\end{equation}
 Therefore in this case the correct prescription is to subtract the IR divergence. This becomes clearer when computing the circular Wilson loop
in the next section. We will then see that the IR divergence is exactly the same and therefore the same subtraction regularizes both cases.

\renewcommand{\theequation}{4.\arabic{equation}}
 \setcounter{equation}{0}

\section{Circular Wilson loop}

First, let us review the string solution dual to the circular Wilson loop \cite{dgo, bcfm}.
We start with the string Nambu action
\begin{equation}
S=\frac{\sqrt{\lambda}}{2 \pi} \int d \sigma d \tau \sqrt{-(\dot{X} X')^2+(\dot{X})^2 (X')^2}
\end{equation}In what follows we use the AdS Euclidian metric in Poicare coordinates
\begin{equation}
ds^2=\frac{1}{z^2}(d r^2+ r^2 d \phi^2 + dz^2 + dx_i^2)+d\Omega^2_5 \label{mp}
\end{equation}
The open string solution corresponding to the circular Wilson loop is
\begin{equation}
z=\sqrt{a^2-r^2},  \quad \quad 0\leq r \leq a, \quad 0\leq \phi < 2\pi
\end{equation}
where the radius of the circle at the boundary of $AdS$ is denoted by $a$. Note that one may rescale away this radius and set it to $1$, but we
keep it arbitrary to check that the physical results do not depend on it. One can translate this solution into embedding coordinates and then
in global coordinates in Minkowski $AdS_5$. The minimal surface ends up on a circle at the boundary of $AdS_5$ and
is diffeomorphic to a disk.

The computation of the area of this solution gives a divergent quantity. One way to regularize it is to
introduce a cutoff $\epsilon$ near the boundary \cite{dgo}, i.e. setting the boundary at $z=\epsilon$, or
equivalently setting the maximum value of $r$ to be $\sqrt{a^2-\epsilon^2}$.
Parametrizing the surface using the coordinates $r,\phi$ one obtains
\begin{equation}
S=\frac{\sqrt{\lambda}}{2 \pi}\int_0^{\sqrt{a^2-\epsilon^2}} d r \int_0^{2 \pi} d \phi \frac{r}{z^2}\sqrt{1+z'^2}=
-\sqrt{\lambda}+\sqrt{\lambda}\frac{a}{\epsilon} \label{tcs}
\end{equation}
The standard procedure as discussed in \cite{dgo} is to set all linear divergencies to zero, so that
area is just
\begin{equation}
S=-\sqrt{\lambda}  \label{jsg}
\end{equation}

In term of the full Euler number (\ref{iis}) one also obtains the regularized area (\ref{jsg}) since the Euler number $\chi$ for a disk is one. The volume and boundary parts of the Euler number in this case are
\begin{equation}
\chi_v= 1- \frac{a}{\epsilon}  \quad  \quad \quad \chi_b= \frac{a}{\epsilon}
\end{equation}
The area in (\ref{tcs}) is proportional to the volume part of the Euler number.
As in the case of the straight string, the natural way to regularize the area is to complete the Euler number by adding a boundary part.

As expected, we see that the physical area (\ref{jsg}) is indeed independent of the radius of the circle. If we now compare the divergent part of the area in (\ref{tcs}) with the divergent part of the area of the straight string with $T=a$, i.e. with (\ref{tst}), we see that they are the same. This is in accord with the expectation that, at the classical level, the linearly divergent part is proportional to the length of the Wilson loop \cite{dgo}.
We should also point out that in the two cases the topologies of the worldsheet are different as well. However, this does not matter at the classical level where only the lengths of the boundaries come into play. To extend this comparison to $1$-loop we will effectively compactify the straight string by choosing periodic boundary conditions in that case as well.

The induced metric on the circular solution is
\begin{equation}
ds^2_2=\frac{r^2}{a^2-r^2}\bigg(\frac{a^2 dr^2}{r^2(a^2-r^2)}+d \phi^2\bigg)
\end{equation}
and the curvature is $R^{(2)}=-2$.
Since at the fluctuation level we prefer to work with the Polyakov action in conformal gauge, let us introduce the coordinate $\sigma$ so that the metric becomes conformaly flat
\begin{equation}
\frac{a d r}{r \sqrt{a^2-r^2}}=d \sigma, \quad \quad ds^2_2= \frac{1}{\sinh^2 \sigma}(d \sigma^2 + d \tau^2)
\end{equation}
which gives the solution in conformal gauge
\begin{equation}
r=\frac{a}{\cosh \sigma}, \quad \quad z=a \tanh \sigma, \quad 0\leq \sigma <\infty, \quad 0\leq \tau \equiv \phi < 2 \pi  \label{ck1}
\end{equation}
where we also have introduced the string coordinate $\tau$.
Note that the cutoff in $z$ at $z=\epsilon$ translates into a cutoff in $\sigma$ at $\epsilon_0$ given by
$\epsilon= a \tanh \epsilon_0$.

\bigskip

Let us also discuss the circular Wilson loop with arbitrary winding $k$ whose corresponding string solution  was discussed in \cite{df}.
It is a simple generalization of (\ref{ck1})
\begin{equation}
r= \frac{a}{\cosh k \sigma}, \quad \quad z= a \tanh k \sigma, \quad \quad \phi=k \tau, \quad \quad 0\leq \tau < 2 \pi, \quad \quad 0\leq \sigma<\infty
\end{equation}
The induced metric on this solution is
\begin{equation}
ds^2_2=\frac{k^2}{\sinh^2 k \sigma}(d \sigma^2 + d\tau^2)
\end{equation}
The computation of the classical action of this solution gives
\begin{equation}
S= - k \sqrt{\lambda}+ \sqrt{\lambda}\frac{a k }{\epsilon} \label{lhd}
\end{equation}
We observe that, as in the previous cases, the linearly divergent part $\frac{1}{\epsilon}$ is proportional to the length of the Wilson loop. The Wilson loop now is a circle of radius $a$ wrapped $k$-times.

Note that now $\sigma$ is to be cut at $\epsilon_0$ given by $\epsilon= a \tanh k \epsilon_0$, so that the physical cutoff in $z$ is always $\epsilon$.
This is different from the straight string case where $z=\sigma$ and so cutoffs in $z$ and $\epsilon$ were the same. For the circular string, however,
the two cutoffs are related but not the same. The physical area is again independent of the radius of the circle. We will see that the $1$-loop
correction is also independent of the radius $a$.

To this end let us also mention the boundary term obtained from the Legendre transform that cancels the linearly divergent term at the classical level. Following the discussion in (\ref{bcd}) in the case of a circular string we obtain
\begin{equation}
\frac{\partial L}{\partial(\partial_{\sigma} z)}= \frac{k \sqrt{\lambda}}{2 a \pi}\frac{1}{\sinh^2 k \sigma}= \frac{k \sqrt{\lambda}}{2 \pi}\frac{a}{\epsilon^2}, \quad \quad \tilde{S}=S - \sqrt{\lambda}\frac{a k}{\epsilon}
\end{equation}
This extra boundary term cancels the linearly divergent term.

\renewcommand{\theequation}{5.\arabic{equation}}
 \setcounter{equation}{0}

\section{Circular Wilson loop solution: $1$-loop correction to the effective action}

\subsection{Winding number $k=1$}

Proceeding in the same way as for the straight string one can obtain the fluctuations Lagrangian near the circular Wilson loop solution. As before let us now proceed by identifying the background metric $g$ with the induced metric $ds^2=\frac{1}{\sinh^2 \sigma}(d\tau^2+d \sigma^2)$.
In this case it is more convenient to
use the metric in polar coordinates (\ref{mp}). Then the worldsheet projections of the vielbein and spin connection are
\begin{equation}
e^a_i=\bigg(\frac{r}{z}\partial_i \phi, \frac{1}{z} \partial_i r, \frac{1}{z} \partial_i x_2, \frac{1}{z} \partial_i x_3, \frac{1}{z} \partial_i z \bigg)  \label{vcc}
\end{equation}
\begin{equation}
\Omega^1_{i  4}=-\frac{1}{z}\partial_i r, \quad \Omega^0_{ i 4}=-\frac{r}{z} \partial_i \phi, \quad \Omega^0_{ i 1}=\partial_i \phi, \quad \Omega^2_{ i 4}=-\frac{1}{z} \partial_i x_2,
\quad \Omega^3_{ i 4}=-\frac{1}{z}\partial_i x_3 \label{spcc}
\end{equation}
On the circular solution these do no depend on the radius of the circle $a$.
The nontrivial covariant derivatives and mass matrices are
\begin{equation}
D_0 \zeta^0= \partial_0 \zeta^0 - \frac{1}{s}\zeta^4+ \zeta^1, \quad D_1 \zeta^0= \partial_1 \zeta^0, \quad D_0 \zeta^1= \partial_0 \zeta^1 - \zeta^0, \quad D_1 \zeta^1=\partial_1 \zeta^1 + \frac{1}{c}\zeta^4
\end{equation}
\begin{equation}
D_0 \zeta^4 = \partial_1 \zeta^4 + \frac{1}{s}\zeta^0, \quad D_1 \zeta^4=\partial_1 \zeta^4 - \frac{1}{c}\zeta^1
\end{equation}
\begin{equation}
\quad X^{ab}=s^2 \text{diag} \bigg(\frac{1}{s^2},\frac{2}{s^2}-\frac{1}{c^2},\frac{2}{s^2},\frac{2}{s^2},\frac{2}{s^2}-\frac{1}{c^2 s^2}\bigg)+\frac{2 s}{c^2 }\delta_1^{(a} \delta_4^{b)}, \quad X^{pq}=0
\end{equation}
where we introduced the notation $s=\sinh \sigma, c=\cosh \sigma$. As in \cite{sy} we rotate $\zeta^1,\zeta^4$ so that the mass matrix becomes diagonal
\begin{equation}
\tilde{X}^{ab}=\text{diag} (1,1,2,2,2)
\end{equation}
After the rotation the only non-trivial covariant derivatives are
\begin{equation}
D_0 \zeta^0 = \partial_0 \zeta^0 - \frac{c}{s}\zeta^1, \quad D_1 \zeta^0= \partial_1 \zeta^0, \quad D_0 \zeta^1=\partial_0 \zeta^1 + \frac{c}{s}\zeta^0, \quad D_1 \zeta^1 = \partial_1 \zeta^1
\end{equation}
where for simplicity of notation we denoted by the same letters fluctuations before and after rotation.
The resulting bosonic fluctuation Lagrangian becomes
\begin{eqnarray}
S=\frac{\sqrt{\lambda}}{4 \pi}\int d \tau d \sigma \frac{1}{s^2} \bigg[s^2 (\partial_0 \zeta^A)^2 &+& s^2 (\partial_1 \zeta^A)^2 +2 \bigg((\zeta^2)^2+
(\zeta^3)^2+(\zeta^4)^2\bigg)+ (s^2+ 2) ((\zeta^0)^2+(\zeta^1)^2)\nonumber\\
&-&2 s c \ \dot{\zeta}^0 \zeta^1+ 2 s c \ \zeta^0 \dot{\zeta}^1\bigg]
\end{eqnarray}
where $A=0,...,9$. As in the case of the straight string the ghost Lagrangian is coupled and it is the same  as the longitudinal fluctuations $\zeta^0,\zeta^1$ \cite{dgt,sy}. Therefore their contributions cancel each other in the partition function.
For the remaining three decoupled transversal modes we need to solve the following spectral problem.
\begin{equation}
Lf= \Lambda f, \quad L=\sinh^2 \sigma (-\partial_0^2-\partial_1^2) + 2 \label{scf2}
\end{equation}
There are also five massless modes from $S^5$, whose spectral problem is the same as in (\ref{scf2}) but with mass equal to zero.

Since we have a circle in $\tau$ we choose periodic boundary conditions in $\tau$. The solutions are of the form $f(\tau,\sigma)=e^{i m \tau} g(\sigma)$ where $m$ is an integer number. $\sigma$ is in the range $\epsilon_0 \leq \sigma < \infty$, and, we take Dirichlet boundary conditions on fluctuations, that is $g(\epsilon_0)=0, g(R)=0$. As in the case of the straight string, for the purpose of computing the determinants, we introduce a large $R$ and at the end we take the limit $R\rightarrow \infty$. This procedure effectively introduces an extra boundary but no effects are left after taking the $R\rightarrow \infty$ limit\footnote{If we choose instead the background metric $g_{ij}$ to be flat, at least naively, effects from the un-physical boundary at $ \sigma=R$ seem to remain.}. We also introduce a small cutoff $\epsilon_0$ to keep track of the divergencies at the boundary of $AdS$. As we already mentioned, we denote the physical cutoff for $z$ to be $\epsilon$. Then $\epsilon=a \tanh \epsilon_0$. In what follows we absorb the radius $a$ into $\epsilon$ as $a$ enters the $1$-loop correction only through $\epsilon$. Since, as we will see, the physical $1$-loop result is independent of $\epsilon$ it means it is independent of $a$ too. This is indeed expected at all orders in loop expansion.  For the particular situation with $m=0$ we take Neumann boundary conditions at $R$ and Dirichlet at $\sigma=\epsilon_0$. The determinant of the operator $L$ can be written as
\begin{equation}
\det L= \prod_{m=-\infty}^{\infty}\det  \bigg(\sinh^2 \sigma (-\partial_1^2 + m^2 )+2\bigg)
\end{equation}
Before dealing with this determinant let us move on to the fermionic part.

\bigskip

\bigskip

Let us now consider the  fermionic Lagrangian for the circular string solution with $g_{ij}$ being the induced metric. In this case it is more convenient to
use the metric in polar coordinates (\ref{mp}). For convenience we choose the vielbiens and spin connection as given in (\ref{vcc},\ref{spcc}).

Note that these were obtained from the corresponding cartesian ones (\ref{vsps}) by performing an angle $\phi$ rotation both, in the space-time indices
$\mu$ and in the tangent space indices $A$. As a consequence, the fermions become antiperiodic.
Alternatively, we could try to work with cartesian vielbiens and periodic fermions. However, the Lagrangian will be explicitely $\phi=\tau$ dependent.
After doing a $\tau$ rotation to get rid of the $\tau$ dependence we arrive at the same fermionic Lagrangian (\ref{hfk}), as below with antiperiodic
fermions. The anti-periodicity of the fermions implies that the quantum number in the $\tau$ direction is half-integer.
We denote half-integer numbers by $r$, while reserving $n,m$ for integers.

 For the circular loop solution we obtain
\begin{equation}
\rho_0=\frac{1}{\sinh \sigma}\Gamma_0, \quad \quad \rho_1= -\frac{1}{\cosh \sigma}\Gamma_1+\frac{1}{\sinh \sigma \cosh \sigma}\Gamma_4
\end{equation}
\begin{equation}
\nabla_0=\partial_0+\frac{1}{2}\Gamma_{01}-\frac{1}{2 \sinh \sigma}\Gamma_{04}, \quad \nabla_1=\partial_1+\frac{1}{2 \cosh \sigma}\Gamma_{14}
\end{equation}
Note that the radius of the circle cancels out in the vielbein and spin connection.
These expressions can be simplified further if we consider the following rotation
\begin{equation}
\theta= e^{-\frac{p}{2}\Gamma_1 \Gamma_4}\Psi, \quad \quad \cos p =\frac{1}{\cosh \sigma}, \quad \quad \sin p=-\tanh \sigma
\end{equation}
Applying this rotation we obtain
\begin{equation}
\rho_0=\frac{1}{\sinh \sigma}\Gamma_0, \quad \quad \rho_1=\frac{1}{\sinh \sigma}\Gamma_4
\end{equation}
\begin{equation}
\nabla_0=\partial_0-\frac{1}{2}\coth \sigma \Gamma_0 \Gamma_4, \quad \quad \nabla_1=\partial_1
\end{equation}
The fermionic Lagrangian becomes
\begin{equation}
L_{2F}=-2 i \sqrt{g}\bar{\Psi}D_F \Psi
\end{equation}
where
\begin{equation}
D_F=- \sinh \sigma \Gamma_0 \partial_0 + \sinh \sigma \Gamma_4 \partial_1-\frac{1}{2}\cosh \sigma \Gamma_4 +i \Gamma_0 \Gamma_4 \label{hfk}
\end{equation}
For small $\sigma$, namely near the boundary, this operator is the same as the corresponding operator for the straight string, as expected.
Using again the same representation for the gamma matrices as for the straight string, we obtain the spectral problem for the fermions
\begin{equation}
L_{F} \theta = \Lambda \theta
\end{equation}
where
\begin{equation}
L_{F}=-\nabla_i \nabla^i + \frac{R^{(2)}}{4}+1=\sinh^2\sigma (-\partial_1^2+r^2)+\frac{3}{4}+\frac{\sinh^2 \sigma}{4}+\Gamma_{04} r  \cosh \sigma \sinh \sigma \label{csf}
\end{equation}
and we have introduced the $\tau$ dependent part $\sim e^{i r \tau}$. Let us recall that $r$ is a half-integer number.

\bigskip

Putting together the bosons and fermions one ends up with the following partition function \cite{dgt}
\begin{equation}
Z=\frac{\det^{8/2} (-\nabla^2 + \frac{R^{(2)}}{4}+1)}{\det^{3/2}(-\nabla^2 +2) \det^{5/2}(-\nabla^2)} \label{part1}
\end{equation}
This has the same form as the partition function for the straight string. However, the induced metric is different so the spectral problems are
actually different. But, for small $\sigma$, the partition functions for the straight and the circular solutions are the same, as the fluctuation
Lagrangians coincide in that limit. Thus, we expect that, with the same boundary conditions, the $\frac{1}{\epsilon}$ terms, which are the dominant
terms in the small $\sigma$ limit, are the same in the two cases. We see below that this is indeed the case, which suggests that we should subtract the
results in the two cases in order to cancel the $\frac{1}{\epsilon}$ divergency. Explicitly in the circular string case the partition function is
\begin{eqnarray}
Z=  \frac{\prod_{r \in \mathbb{Z}+\frac{1}{2}} \ \det^{4/2}[s^2 (-\partial_1^2 + r ^2)+ \frac{3}{4}+ \frac{s^2}{4}+ r s \ c] \ \det^{4/2}[s^2 (-\partial_1^2 + r^2)+ \frac{3}{4}+\frac{s^2}{4}- r s \ c] }{\prod_{m \in \mathbb{Z}}\det^{3/2}[s^2 (-\partial_1^2 + m^2)+2]
\ \det^{5/2}[s^2 (-\partial_1^2 + m^2)]}
\end{eqnarray}
As in the case of the straight string the factor $s^2$ can be scaled away when computing the above ratios of determinants since the initial value problems are the same. The inconvenience we face here is that the bosonic sum/product is over integers while the fermionic one is over half integers.
Let us rewrite  the fermionic products as products over integers by performing shifts in the summation indices. As in \cite{fpt} we perform these shifts in a `supersymmetric way'. In order to work with finite quantities let us consider the ratio of the fermionic determinants by $\det  (-\partial_1^2)$.
Consider then
\begin{equation}
\sum_{r\in \mathbb{Z}+\frac{1}{2}} \omega_r = \sum_{r\in \mathbb{Z}+\frac{1}{2}} \ln \frac{\det (-\partial_1^2 + r^2 + \frac{1}{4}+\frac{3}{4 \sinh^2 \sigma}+ r \coth \sigma)}{\det (-\partial_1^2)}
\end{equation}
The fermionic determinants that we need to compute can be regularized by introducing a suppressing exponential factor:
\begin{equation}
\sum_{r\in \mathbb{Z}+\frac{1}{2}} \omega_r + \sum_{r\in \mathbb{Z}+\frac{1}{2}} \omega_{-r}\ \ \ \rightarrow
\sum_{r\in \mathbb{Z}+\frac{1}{2}} e^{- \mu |r|} \omega_r + \sum_{r\in \mathbb{Z}+\frac{1}{2}} e^{- \mu |r|} \omega_{-r} \label{isd}
\end{equation}
Considering the shifts $r=m-\frac{1}{2}$ in the first sum and $r=m+\frac{1}{2}$  in the second sum, with $m\in \mathbb{Z}$, we obtain for the above sum
\begin{eqnarray}
\sum_{r\in \mathbb{Z}+\frac{1}{2}} \omega_r + \sum_{r\in \mathbb{Z}+\frac{1}{2}} \omega_{-r}&= &\sum_{m \in \mathbb{Z}} e^{-\mu |m|}(\omega_{m-\frac{1}{2}}+\omega_{-m-\frac{1}{2}})\\
&+& \sum_{m \in \mathbb{Z}}\omega_{m-\frac{1}{2}} (e^{-\mu |m-\frac{1}{2}|}-e^{-\mu |m|})+ \sum_{m \in \mathbb{Z}}\omega_{-m-\frac{1}{2}} (e^{-\mu |m+\frac{1}{2}|}-e^{-\mu |m|})\nonumber
\end{eqnarray}
We take the limit $\mu \rightarrow 0$. The sums in the second line above can be evaluated giving a finite result. The sum in the first line is divergent but its divergency is the same as the original sum. The sum in the second line is even in $m$ and taking the $\mu\rightarrow 0$ limit we obtain
\begin{equation}
\sum_{r\in \mathbb{Z}+\frac{1}{2}} \omega_r + \sum_{r\in \mathbb{Z}+\frac{1}{2}} \omega_{-r}=\sum_{m \in \mathbb{Z}} (\omega_{m-\frac{1}{2}}+\omega_{-m-\frac{1}{2}})+ \mu \sum_{m=1}^{\infty} e^{-\mu m}(\omega_{m-\frac{1}{2}}-\omega_{-m-\frac{1}{2}}) \label{uas}
\end{equation}
To evaluate the second sum in the above expression we need the ratio of $\omega_{m-\frac{1}{2}}$ and $\omega_{-m-\frac{1}{2}}$. Let us compute them with the determinant ratio method employed in this paper. We need the initial value solution for the equation
\begin{equation}
 \bigg[-\partial_1^2+r^2+\frac{1}{4}+\frac{3}{4 \sinh^2 \sigma}+ r \coth \sigma\bigg]\theta=0, \quad \quad \theta(\epsilon)=0, \quad \quad \theta'(\epsilon)=1 \label{stf1}
 \end{equation}
The solution is
\begin{equation}
\theta(\sigma)= \frac{1}{4r ^2-1}\frac{1}{\sqrt{\sinh \epsilon_0}}\bigg[(2 r \tanh \sigma-1) \frac{\cosh \sigma}{\sqrt{\sinh \sigma}}e^{ r(\sigma-\epsilon_0)}+(1-2 r \tanh \epsilon_0) \frac{\cosh \epsilon_0}{\sqrt{\sinh \sigma}}e^{-r (\sigma-\epsilon_0)}\bigg]
\end{equation}
For the free operator we need
\begin{equation}
(-\partial_1^2) f =0, \quad \quad f(\epsilon)=0, \quad \quad f'(\epsilon)=1
\end{equation}
with solution
\begin{equation}
f(\sigma)=\sigma-\epsilon
\end{equation}
Taking the solutions in the large $\sigma$ limit we obtain the ratios of determinants needed in (\ref{uas})
\begin{equation}
\omega_{m-\frac{1}{2}} = \ln \frac{1}{2 m \sqrt{2 \sinh \epsilon_0}  } \frac{e^{m R}}{R} e^{-(m-\frac{1}{2})\epsilon_0}
\end{equation}
\begin{equation}
\omega_{-m-\frac{1}{2}} = \ln \frac{1}{2 m \sqrt{2 \sinh \epsilon_0}  }\frac{1+ (2m+1) \tanh \epsilon_0}{m+1}\cosh \epsilon_0 \frac{e^{m R}}{R}e^{-(m+\frac{1}{2})\epsilon_0}
\end{equation}
Using the relation between $\epsilon$ and $\epsilon_0$, after performing the second sum in (\ref{uas}) we obtain the cutoff independent result
\begin{equation}
\sum_{r\in \mathbb{Z}+\frac{1}{2}} \omega_r + \sum_{r\in \mathbb{Z}+\frac{1}{2}} \omega_{-r}=\sum_{m \in \mathbb{Z}} (\omega_{m-\frac{1}{2}}+\omega_{-m-\frac{1}{2}})+ \ln \frac{1+ \epsilon}{2 \epsilon}
\end{equation}
where the last term is to be expanded in small $\epsilon$. We therefore conclude that the sum over half-integers in the fermionic determinants is transformed in a sum over integer as
\begin{eqnarray}
&&\prod_{r\in \mathbb{Z}+\frac{1}{2}} \frac{\det[-\partial_1^2 + r ^2+ \frac{3}{4 s^2}+ \frac{1}{4}+ r  \frac{c}{s}]}{\det[-\partial_1^2]}\frac{\det[-\partial_1^2 + r ^2+ \frac{3}{4 s^2}+ \frac{1}{4}- r  \frac{c}{s}]}{\det[-\partial_1^2]} \label{wdj}\\
&=&\frac{1}{2 \epsilon}\prod_{m\in \mathbb{Z}} \frac{\det[-\partial_1^2 + (m-\frac{1}{2}) ^2+ \frac{3}{4 s^2}+ \frac{1}{4}+ (m-\frac{1}{2})  \frac{c}{s}]}{\det[-\partial_1^2]}\frac{\det[-\partial_1^2 + (m+\frac{1}{2}) ^2+ \frac{3}{4 s^2}+ \frac{1}{4}- (m+\frac{1}{2})  \frac{c}{s}]}{\det[-\partial_1^2]}\nonumber
\end{eqnarray}

The $1$-loop effective action can therefore be written as
\begin{equation}
\Gamma_1= \frac{1}{2}(4 \ln 2 + 4 \ln \epsilon + \sum_{m=-\infty}^{\infty} \ln P_m) \label{pfg}
\end{equation}
where
\begin{equation}
P_m= \frac{\bigg[\det[-\partial_1^2+m^2 +\frac{2}{s^2}]\bigg]^3 \bigg[\det[-\partial_1^2+m^2]\bigg]^5}{\bigg[\det[-\partial_1^2+(m-\frac{1}{2})^2 +\frac{1}{4}+\frac{3}{4 s^2}+ (m-\frac{1}{2}) \frac{c}{s}\bigg]^4 \bigg[\det[-\partial_1^2+(m+\frac{1}{2})^2 +\frac{1}{4}+\frac{3}{4 s^2}- (m+\frac{1}{2}) \frac{c}{s}\bigg]^4}
\end{equation}
Note that the symmetry under $m\rightarrow -m$ is preserved by the supersymmetric shifts performed, so we can restrict ourselves to $m>0$.

We now compute the ratio of determinants. Let us focus first in the case with $m \neq 0$. The initial value problem for the transverse bosons that we need is
\begin{equation}
-g''+\bigg(m^2+\frac{2}{\sinh^2 \sigma}\bigg) g = 0
\end{equation}
with the initial conditions
\begin{equation}
g(\epsilon_0)=0, \quad g(\epsilon_0)=1
\end{equation}
The initial value solution is
\begin{equation}
g(\sigma)=\frac{1}{2 m (m^2-1)}\bigg[(m+\coth \epsilon_0)(m-\coth \sigma)e^{m (\sigma-\epsilon_0)}-(m-\coth \epsilon_0)(m+\coth \sigma)e^{-m (\sigma-\epsilon_0)}\bigg]
\end{equation}
This is valid for $m\neq 1$. For $m=1$ the solution is
\begin{equation}
g(\sigma)=-\frac{1}{4} \frac{1}{\sinh \epsilon_0 \sinh \sigma}(2 (\sigma-\epsilon_0)+ \sinh 2 \epsilon_0 - \sinh 2 \sigma)
\end{equation}
We also need the solution for the free massive bosons
\begin{equation}
-g''+ m^2 g = 0, \quad \quad g(\epsilon_0)=0, \quad g(\epsilon_0)=1
\end{equation}
with solution
\begin{equation}
g(\sigma)=\frac{1}{m}\sinh (m (\sigma-\epsilon_0))
\end{equation}
For the fermionic operators we already wrote the solution above. The ratios of determinants needed are (for $m\neq 0$)

\begin{equation}
\frac{\det[-\partial_1^2+m^2 + \frac{2}{\sinh^2 \sigma}]}{\det[-\partial_1^2 +(m-\frac{1}{2})^2 +\frac{1}{4}+\frac{3}{4 \sinh^2 \sigma}+(m-\frac{1}{2}) \coth \sigma]}=\sqrt{2 \sinh \epsilon_0} \frac{m+ \coth \epsilon_0}{m+1}e^{-\frac{\epsilon_0}{2}} \ ,
\end{equation}
\begin{equation}
\frac{\det[-\partial_1^2+m^2]}{\det[-\partial_1^2 +(m-\frac{1}{2})^2 +\frac{1}{4}+\frac{3}{4 \sinh^2 \sigma}+(m-\frac{1}{2}) \coth \sigma]}=\sqrt{2 \sinh \epsilon_0} e^{-\frac{\epsilon_0}{2}} \ ,
\end{equation}
\begin{equation}
\frac{\det[-\partial_1^2+m^2 ]}{\det[-\partial_1^2 +(m+\frac{1}{2})^2 +\frac{1}{4}+\frac{3}{4 \sinh^2 \sigma}- (m+\frac{1}{2}) \coth \sigma]}=\frac{\sqrt{2 \sinh \epsilon_0}}{\cosh \epsilon_0}\frac{m+1}{1+ (2 m+1)\tanh \epsilon_0}e^{\frac{\epsilon_0}{2}}
\end{equation}

Changing $\epsilon_0$ to $\epsilon$ and putting these ratios together we obtain
\begin{equation}
P_m=\frac{(m+ \frac{1}{\epsilon})^3(m+1)}{(m+ \frac{1}{2}+\frac{1}{2 \epsilon})^4} \label{fcss}
\end{equation}
It is a nontrivial check that this result is independent of the regulator $R$. This had to be the case since we introduced a non-physical boundary
at $\sigma=R$ just to regulate the determinants.
We observe that, as expected, the series does not converge; there is a logarithmic divergence which as we will see below is proportional to the volume part of the Euler characteristic, and the coefficient is the same as in the case of a straight string. Since the sum is divergent we introduce a large cutoff $\Lambda$. Before doing that let us also compute the ratios of determinants for $m=0$.

As we discussed already, for $m=0$ we take Neumann boundary conditions in $\sigma$ at $\sigma=R$.
For the transversal modes we need
\begin{equation}
\bigg(-\partial_1^2 + \frac{2}{\sinh^2 \sigma}\bigg)g =0 , \quad g(\epsilon_0)=0, \quad g'(\epsilon_0)=1
\end{equation}
with solution
\begin{equation}
g(\sigma)= \coth \sigma + \coth \epsilon_0 [(\sigma-\epsilon_0) \coth \sigma -1]
\end{equation}
For large $\sigma$ this becomes simple
\begin{equation}
g = \sigma \coth \epsilon_0, \quad \quad g'= \coth \epsilon_0
\end{equation}
For the fermions we need to solve
\begin{equation}
\bigg(-\partial_1^2 + \frac{1}{2}+ \frac{3}{4 \sinh^2 \sigma}-\frac{1}{2}\coth \sigma \bigg)\theta=0, \quad \quad \theta(\epsilon_0)=0, \quad \quad \theta'(\epsilon_0)=1
\end{equation}
The solution is
\begin{equation}
\theta(\sigma)= -\frac{1}{2}\sqrt{1- \coth \epsilon_0}\sqrt{1- \coth \sigma}[ e^{(\sigma + \epsilon_0)} (\sigma-\epsilon_0) + \sinh (\epsilon_0-\sigma)]
\end{equation}
which in the large $\sigma$ limit becomes
\begin{equation}
\theta= \sigma \frac{e^{\frac{\epsilon_0}{2}}}{\sqrt{2 \sinh \epsilon_0}}, \quad \quad \theta' =\frac{e^{\frac{\epsilon_0}{2}}}{\sqrt{2 \sinh \epsilon_0}}
\end{equation}
Note that none of the derivatives of the above solutions vanish at $\epsilon_0=0$, so there are no zero modes present. Taking the derivatives of these function at large $\sigma$ as appropriate for Neumann boundary conditions (see Appendix A) we obtain
\begin{equation}
P_0= \frac{16 \epsilon}{(1+ \epsilon)^4} \ \ \Rightarrow \ \ \ \ln P_0 \simeq 4 \ln 2 + \ln \epsilon, \ \ (\epsilon\rightarrow 0)
\end{equation}
where we have expanded in small $\epsilon$. We observe that this is in fact the same as $P_m$ from (\ref{fcss}) with $m=0$.

Let us plug (\ref{fcss}) into (\ref{pfg}) and express the sum in terms of gamma functions introducing again a dimensionless cutoff $\Lambda$ in the summation indexes $m$. We obtain the result
\begin{equation}
\Gamma_1= \frac{1}{\epsilon}(1+ \ln \frac{\epsilon}{4})+ \frac{1}{\epsilon}\ln \Lambda - \ln \Lambda +\frac{1}{2}\ln (2 \pi) \label{gfd}
\end{equation}

Having in view the volume part of the Euler number (here $a$ is absorbed in $\epsilon$)
\begin{equation}
\chi_v=\frac{1}{4\pi}\int_0^{2 \pi} d \tau \int_{\epsilon_0}^\infty d \sigma \frac{-2}{\sinh^2 \sigma}=1- \coth \epsilon_0=1-\frac{1}{\epsilon}
\end{equation}
 Again, the UV divergent part in the $1$-loop effective action is proportional to the volume part of the Euler characteristic,
$\Gamma_1 \rightarrow -\chi_v \ln \Lambda$, with precisely the same coefficient in front as in the case of a straight string.
This is of course expected for the consistency of the method. For any Wilson loop solution, working with the induced metric and
with genuine $2d$ fermions, one should obtain the same UV divergency factor form the $1$-loop fluctuation determinants. As in the case
of the straight string, the proper regularization should complete the Euler number in the UV divergency, which of course should be
canceled by a corresponding factor from the measure. Any such completion of the Euler number with a boundary term goes like $\frac{1}{\epsilon}$, thus no finite part can remain.

 For completeness let us restore the radius of the circle $\epsilon \rightarrow \frac{\epsilon}{a}$, so that $1$-loop effective action becomes
\begin{equation}
\Gamma_1= \frac{a}{\epsilon}(1+ \ln \frac{\epsilon}{4 a})  + \frac{1}{2}\ln ( 2 \pi)  \label{gfdd}
\end{equation}

 As expected, the finite part of the $1$-loop effective action is independent of the radius of the circle.
The ($\epsilon\rightarrow 0$) divergent part is the same as the one for a straight string of length $T=a$. If we subtract both the result is finite:
\begin{equation}
\Gamma_1=  \frac{1}{2}\ln ( 2 \pi)
\end{equation}
In addition to the part of $1$-loop effective action computed above, there is a contributing numerical factor from the normalization of the zero modes.
As pointed out in \cite{dg}, in the case of a disk there are three normalizable zero modes\footnote{The three zero modes come from the residual
$SL(2,\mathbb{R})$ symmetry after fixing the metric to be the induced metric. There are no Teichmuller parameters for a disk.
The group $SL(2,\mathbb{R})$ is noncompact, so its volume, which appears in the measure, should be regularized as in \cite{a,l}.}.
 The contribution from the normalization of zero modes has the form $c \lambda^{-3/4}$.
The precise numerical factor $c$ is ambiguous.  Assuming, for the moment, that $c=1$, we observe that the result in (\ref{gfdd})
and the gauge theory expectation (\ref{gc}) differ by $\ln 2$. Equivalently the partition function that we obtain is half of the
expected partition function from gauge theory, i.e. $Z= \frac{1}{2}<W>$. To have $<W>=Z$, as predicted by $AdS/CFT$, one needs to have $c=2$. It would be interesting to obtain this factor $c$ in the string partition function.

\subsection{Arbitrary winding number $k$}

Let us now generalize the above discussion for the string solution with arbitrary winding $k$. The extension is straightforward, and the fluctuation Lagrangian becomes
\begin{eqnarray}
S=\frac{\sqrt{\lambda}}{4 \pi}\int d \tau d \sigma \frac{1}{s^2} \bigg[s^2 (\partial_0 \zeta^A)^2 &+& s^2 (\partial_1 \zeta^A)^2 +2 k^2 \bigg((\zeta^2)^2+
(\zeta^3)^2+(\zeta^4)^2\bigg)+ k^2(s^2+ 2) ((\zeta^0)^2+(\zeta^1)^2)\nonumber\\
&-&2k \  s \ c \ \dot{\zeta}^0 \zeta^1+ 2 k \ s \ c \ \zeta^0 \dot{\zeta}^1\bigg]
\end{eqnarray}
where now $s=\sinh k \sigma$, $c=\cosh k \sigma$. As in  the $k=1$ case one can show that the ghost and longitudinal modes Lagrangians are the same, so their contributions cancel in the partition function. For the fermions we find that the relevant quadratic fermionic operator is
\begin{equation}
D_F^2= \sinh^2 k \sigma (-\partial_1^2+r^2)+\frac{3 k^2}{4}+\frac{k^2}{4}\sinh^2 k \sigma+\Gamma_{04} r k  \cosh k \sigma \sinh k \sigma
\end{equation}
The shifts that we do to transform the summation over half-integer into integer summation are $r=m-\frac{k}{2}$ and $r=m+\frac{k}{2}$ in the two sums in (\ref{isd}). Proceeding like in the $k=1$ case we obtain a factor of $(\frac{1}{2 \epsilon})^k$ in an expression similar to (\ref{wdj}).

For the transverse bosons we need the initial value solution for the equation
\begin{equation}
-g'' +\bigg(m^2 + \frac{2 k^2}{\sinh^2 k \sigma}\bigg)g=0
\end{equation}
with solution for $m \neq k$
\begin{equation}
g(\sigma)=\frac{1}{2 m (m^2-k^2)}\bigg[(m+k \coth k \epsilon_0)(m-k \coth k \sigma)e^{m (\sigma-\epsilon_0)}-(m-k \coth k \epsilon_0)(m+k \coth k \sigma)e^{-m (\sigma-\epsilon_0)}\bigg]
\end{equation}
while the solution for $m=k$ is
\begin{equation}
g(\sigma)=-\frac{1}{4 k} \frac{1}{\sinh k \epsilon_0 \sinh k \sigma}[2 k (\sigma-\epsilon_0)+ \sinh 2 k \epsilon_0 - \sinh 2 k \sigma]
\end{equation}

For the fermionic operator we need
\begin{equation}
 \bigg[-\partial_1^2+r^2+\frac{k^2}{4}+\frac{3 k^2}{4 \sinh^2 k\sigma}+ r k \coth k\sigma\bigg]\theta=0, \label{stff2}
 \end{equation}
The solution is
\begin{eqnarray}
\theta(\sigma)&=& \frac{1}{4r^2-k^2}\frac{1}{\sqrt{\sinh k \epsilon_0}}\bigg[(2 r \tanh k \sigma-k) \frac{\cosh k \sigma}{\sqrt{\sinh k \sigma}}e^{r (\sigma-\epsilon_0)}\nonumber\\
&+&(k- 2 r \tanh k \epsilon_0) \frac{\cosh k \epsilon_0}{\sqrt{\sinh k \sigma}}e^{- r(\sigma-\epsilon_0)}\bigg]
\end{eqnarray}
Computing the relevant determinants we obtain

\begin{equation}
\frac{\det[-\partial_1^2+m^2 + \frac{2 k^2}{\sinh^2 k \sigma}]}{\det[-\partial_1^2 +(m-\frac{k}{2})^2 +\frac{k^2}{4}+\frac{3 k^2}{4 \sinh^2 k \sigma}+(m-\frac{k}{2}) k \coth k \sigma]}=\sqrt{2 \sinh k \epsilon_0} \frac{m+ k \coth k \epsilon_0}{m+k}e^{-\frac{k \epsilon_0}{2}} \ ,
\end{equation}
\begin{equation}
\frac{\det[-\partial_1^2+m^2 ]}{\det[-\partial_1^2 +(m-\frac{k}{2})^2 +\frac{k^2}{4}+\frac{3 k^2}{4 \sinh^2 k \sigma}+(m-\frac{k}{2}) k \coth k \sigma]}=\sqrt{2 \sinh k \epsilon_0} e^{-k \frac{\epsilon_0}{2}} \ ,
\end{equation}
\begin{equation}
\frac{\det[-\partial_1^2+m^2 ]}{\det[-\partial_1^2 +(m+\frac{k}{2})^2 +\frac{1}{4}+\frac{3}{4 \sinh^2 k \sigma}- (m+\frac{k}{2}) \coth k \sigma]}=\frac{\sqrt{2 \sinh k \epsilon_0}}{\cosh k \epsilon_0}\frac{m+k}{k+ (2 m+k)\tanh k \epsilon_0}e^{\frac{k \epsilon_0}{2}}
\end{equation}
Note that these ratios of determinants are precisely those at $k=1$ with the rescaling $m \rightarrow \frac{m}{k}$. This is expected since the solution with arbitrary $k$ can be transformed to the one with $k=1$ with rescaling of coordinates $\tau \rightarrow \frac{\tau}{k}$, $\sigma \rightarrow \frac{\sigma}{k}$. At the level of the classical action the effect of this rescaling is a factor of $k$, as we have seen, but at $1$-loop level this is no longer true even at large $k$.

Putting all together the $1$-loop effective action is
\begin{equation}
\Gamma_1= \frac{1}{2}(4 k \ln 2 + 4 k \ln \epsilon + 4 \ln 2 + \ln \epsilon + 2 \sum_{m=1}^{\infty} \ln P_m)
\end{equation}
where the first two terms come from the $r$-shifting while the next two from $P_0$. Also, $P_m$ now is
\begin{equation}
P_m=\frac{(m+ \frac{k}{\epsilon})^3(m+k)}{(m+ \frac{k}{2}+\frac{k}{2 \epsilon})^4} \label{fcssf}
\end{equation}

Doing the sum with a cutoff $\Lambda$ we obtain (we also restore the radius of the circle)
\begin{equation}
\Gamma_1= \frac{a k}{\epsilon}(1+ \ln \frac{\epsilon}{4 a k})- \chi_v \ln \Lambda +\frac{1}{2}[\ln (2 \pi)+ (4 k+1) \ln k-2 \ln \Gamma(1+k)]  \label{gfa}
\end{equation}
Here again the UV divergent part is proportional to the volume part of the Euler number (here $\chi_v$ is proportional to $k$), and it should be canceled by the measure. The $\ln \epsilon$ divergencies (not $\frac{1}{\epsilon}\ln \epsilon$) cancel by themselves non-trivially. To get rid of the remaining $\frac{1}{\epsilon}$ divergency we again subtract the straight string now with length $T= a k$.   As mentioned before we cut a small region at the center of the disk where the induced metric is singular.
We do not expect this to introduce any problems but more analysis is needed to be completely sure that the result is compatible or incompatible with the field theory
prediction.

Let us finish this section by considering the finite result (\ref{gfa}) for large $k$
\begin{equation}
\Gamma_1=  k \ln k + k + O(\frac{1}{k})  \label{hef}
\end{equation}
The same result can be obtained if one uses the Euler-Maclaurin formula that transforms the sum into an integral plus a remaining sum.
For large $k$ this can be viewed as a decompactifying  limit of the solution with $k=1$. After performing the above rescalings of $\tau$
and $\sigma$, the new $\tau$ runs from $0$ to $2 \pi k$, which is a large interval for large $k$. In contrast to the situation
in \cite{ftt,rtt} where no IR divergencies were present, and in the decompactifying limit the volume factorizes also at $1$-loop,
in our case this is not true as we obtain the result in (\ref{hef}) at large $k$.


\bigskip

\section*{Acknowledgments }
We are  grateful to O. Alvarez, N. Drukker, K.~Kirsten, A.~J.~McKane, and V. Pestun
for useful discussions. We are especially grateful to A. Tseytlin for discussions and comments on earlier versions of this paper.
This work was supported in part by NSF under grant PHY-0653357.

\renewcommand{\theequation}{A.\arabic{equation}}
\renewcommand{\thesection}{A}
 \setcounter{equation}{0}
\setcounter{section}{1} \setcounter{subsection}{0}

 \section*{Appendix A:
Ratio of determinants of second-order differential operators}

In this appendix we briefly review the method of computing the ratio of determinants that we employ in this paper. Let us consider the following
second-order differential operators defined on an interval $\sigma\in [a,b]$
\begin{equation}
L= - P_0(\sigma) \frac{d^2}{d\sigma^2}+ P_1(\sigma) \frac{d}{d\sigma}+P_2(\sigma), \quad \hat{L}=- P_0(\sigma) \frac{d^2}{d\sigma^2}+ \hat{P}_1(\sigma) \frac{d}{d\sigma}+\hat{P}_2(\sigma) \label{oef}
\end{equation}
Note that for what follows it is important that the functions multiplying the second derivatives are the same for the two operators.
In what follows the inner product of eigenfunctions is defined with trivial measure\footnote{In our case of interest in this paper the norm of fluctuations is not trivial. However, as we integrate over fluctuations we can put this norm back into the trivial form by introducing a local determinant. This local determinant does not change the finite part of the result, so we will ignore such determinants.} on $\sigma$. Assuming Dirichlet boundary conditions at the boundary, it was shown in \cite{f} (see also \cite{mt} for a review) that the ratio of determinants of these operators is
\begin{eqnarray}
\frac{\det L}{\det \hat L}=\frac{e^{-\frac{1}{2}\int_a^b d\sigma P_1(\sigma) P_0^{-1}(\sigma)}}{e^{-\frac{1}{2}\int_a^b d\sigma \hat{P}_1(\sigma) P_0^{-1}(\sigma)}}\frac{\psi(b)}{\hat{\psi}(b)}  \label{genr}
\end{eqnarray}
where $\psi$ and $\hat{\psi}$ are solutions of the initial value problems
\begin{equation}
L \psi = 0, \quad \quad \hat{L}\hat{\psi}=0, \quad \psi(a)=\hat{\psi}(a)=0, \quad \psi'(a)=\hat{\psi}'(a)=1  \label{pow}
\end{equation}
This formula is valid when the operators $L$ and $\hat{L}$ do not have zero modes. If zero modes are present one needs corrections as shown in \cite{mt}.
Also, this formula can be generalized for any boundary conditions and for systems of differential operators \cite{km1,km2,mt}. In particular if one needs Dirichlet boundary conditions at $\sigma=a$ but Neumann boundary conditions at $\sigma=b$ one needs to replace the functions by their derivatives in the right hand side in (\ref{genr}), i.e. $\frac{\psi'(b)}{\hat{\psi}'(b)}$.

The particular case of interest to us is when $P_1(\sigma)=\hat{P}_1(\sigma)=0$. Let us further focus on this situation as this is the type of operators we are interested in. Then (\ref{genr}) reduces to
\begin{eqnarray}
\frac{\det L}{\det \hat L}=\frac{\psi(b)}{\hat{\psi}(b)}  \label{genr1}
\end{eqnarray}
This formula (\ref{genr1}) for computing ratio of determinants is valid for intervals $[a,b]$ in which the function $P_0$ does not vanish at any point.
This is indeed the case considered in this paper for the interval $[\epsilon,R]$. However, for $\epsilon$ equals zero,  $P_0$ in our case ($P_0= \sigma^2$
or $P_0= \sinh^2 \sigma$) does vanish. For that reason we always take $\epsilon$ small but non-zero. The divergences that appear in the limit
$\epsilon\rightarrow 0$ are then treated by subtracting a reference solution which in this case is the straight string. This is the same as is
done for the classical area which is also divergent when $\epsilon\rightarrow 0$.

Let us observe that the initial problem solutions $\psi$ and $\hat{\psi}$ are also solutions for the rescaled operators
\begin{equation}
\mathfrak{L}=-\frac{d^2}{d \sigma^2}+ \frac{P_2(\sigma)}{P_0(\sigma)}, \quad \mathfrak{\hat{L}}=-\frac{d^2}{d \sigma^2}+ \frac{\hat{P}_2(\sigma)}{P_0(\sigma)}
\end{equation}
Therefore, the ratio of rescaled and initial operators is the same
\begin{equation}
\frac{\det L}{\det \hat{L}}=\frac{\det \mathfrak{L}}{\det \hat{\mathfrak{L}}}=\frac{\psi(b)}{\hat{\psi(b)}}  \label{yef}
\end{equation}
This is the formula that we used to find the ratio between the determinants of bosonic and fermionic operators. Again this relationship does not include any boundary term that may arise at $\epsilon=0$.

\bigskip

\bigskip

 In the reminder of this appendix we check that the ratio of determinants obtained by the method described above is the same as the ratio computed in the standard way. This comparison can be carried out explicitly for the constant masses determinants where one can compute the spectrum exactly.
Let us consider computing the ratio
\begin{equation}
K=\frac{\det[-\partial_1^2+ \omega^2]}{\det[-\partial_1^2]}
\end{equation}
we take Dirichlet boundary conditions in $\sigma$ with $\sigma \in [0,R]$.
First let us compute $K$ with the method used in this paper
\begin{equation}
K= \frac{\chi(R)}{\chi_0 (R)}
\end{equation}
where $\chi$ and $\chi_0$ satisfy the initial value problem
\begin{equation}
\chi(0)=0, \quad \quad \chi'(0)=1, \quad \quad \chi_0(0)=1, \quad \quad \chi_0'(0)=1
\end{equation}
The solutions are
\begin{equation}
\chi=\frac{1}{\omega}\sinh \omega \sigma, \quad \quad \chi_0= \sigma
\end{equation}
Therefore we obtain (after taking a $\ln$ which is usually what one needs)
\begin{equation}
\ln K= \ln \frac{\sinh(\omega R)}{\omega R}
\end{equation}
This simple computation works for any finite $R$. In the large $R$ limit one obtains
\begin{equation}
\ln K= \omega R - \ln (\omega R) \sim \omega R  \label{largeR}
\end{equation}

\bigskip

\bigskip

Let us now compute the same ratio but with the usual method. The spectrum of the operator
\begin{equation}
(-\partial_1^2 + \omega ) \psi = \lambda \psi
\end{equation}
is given by $\lambda_n= \omega^2 + \frac{\pi^2 n^2}{R^2}$ where $n=1,2,3...$. Then the computation of the determinant is therefore given by
\begin{equation}
\det[-\partial_1^2 + \omega^2] = \prod_{n=1}^{\infty} \bigg(\omega^2+\frac{\pi^2 n^2}{R^2}\bigg)
\end{equation}
Taking again also the same determinant for $\omega=0$ we obtain
\begin{equation}
\ln \frac{\det [-\partial_1^2+\omega^2]}{\det [-\partial_1^2]}=\sum_{n=1}^\infty \bigg[\ln \bigg(\omega^2+ \frac{\pi^2 n^2}{R^2}\bigg)-\ln \frac{\pi^2 n^2}{R^2}\bigg]
\end{equation}
One can compute this sum taking a derivative in respect to $\omega$ and the integrating back after doing the sum and the result is
\begin{equation}
\ln K= \ln \frac{\sinh( \omega R)}{\omega R} \label{secm}
\end{equation}
We see that the result is the same as the one obtained before by the method of wavefunctions. The advantage of the latter method is that it does not require knowing all eigenvalues but only the solution of the initial value problem, which one can often obtain even for complicated potentials.

Let us look at what happens directly when the interval is infinite, i.e. $\sigma\in [0,\infty)$. In this case the spectrum of the operator
\begin{equation}
(-\partial_1^2 + \omega^2) \psi = \lambda \psi
\end{equation}
is continuous $\lambda=\omega^2+ k^2$, where $k> 0$ is a continuous parameter. Here we have Dirichlet boundary condition only at $x=0$.
One needs then to compute
\begin{equation}
\ln K= \ln \prod_{k>0} \frac{k^2+\omega^2}{k^2}
\end{equation}
where the product is over the continuum values of $k> 0$. To define this product ones takes the discrete version introducing a finite volume, and then converts the resulting sum into an integral. Therefore
\begin{equation}
\ln K= \ln \prod_{k>0} \frac{k^2+\omega^2}{k^2} = \frac{R}{\pi}\int_0^\infty dk \ln \frac{k^2+\omega^2}{k^2}= R \omega
\end{equation}
where the above result is valid for large $R$. This is of course exactly what one obtains taking large $R$ limit in (\ref{secm}), or in (\ref{largeR}).

In this appendix we showed that the ratio of the above determinants for free massive operators is the same if computed by the usual method or by the wavefunction method for any $R$. Whether in the strict $R\rightarrow \infty$ limit the result makes or not sense is another matter. In the strict limit the ratio of determinants considered above appears to diverge even though none of the operators has zero eigenvalues. Therefore, in general one needs to consider the ratio of determinants in the strict $R\rightarrow\infty$ with caution. The result to be fully trusted is for the ratios of determinants which are finite (not zero) in the $R\rightarrow \infty$ limit. This is the case in this paper for the straight and circular Wilson loop string solutions where the results are $R$-independent for large $R$. Another situation where the $R$ dependence goes away from the ratio of two determinants was considered in \cite{mt}.


\end{document}